\documentclass[letterpaper,twocolumn,10pt]{article}
\usepackage{ligroup}
\usepackage{xspace} 
\usepackage{pifont}
\usepackage{placeins}
\usepackage{amssymb} 
\usepackage[absolute]{textpos}

\AtBeginDocument{
  }

\usepackage{tikz}
\usepackage{amsmath}
\newcommand*\filledcircled[2][\normalsize]{%
  \tikz[baseline=(char.base)]{
    \node[shape=circle,fill,inner sep=0.5pt] (char) {#1\textcolor{white}{#2}};}}

\usepackage[hyphens]{xurl} 
\usepackage{hyperref}
\hypersetup{
  colorlinks,
  urlcolor={orange!70!red}
}
\usepackage[most]{tcolorbox}
\usepackage{filecontents}
\usepackage[utf8]{inputenc}
\usepackage{amsmath} 
\usepackage{amsfonts} 
\usepackage{booktabs} 
\usepackage{algorithm} 
\usepackage{algpseudocode} 
\usepackage{graphicx}    
\usepackage{subcaption}  
\usepackage{caption}     
\usepackage{multirow}   
\usepackage{booktabs}  
\usepackage{xcolor}     
\usepackage{colortbl} 
\newcommand{\ourframework}{\textit{VidLeaks}}

\begin{document}

\begin{textblock}{15}(1.5,1)
To appear in the Proceedings of the 35th USENIX Security Symposium (USENIX Security ’26), August 12--14, 2026.
\end{textblock}


\date{}

\title{\Large \bf VidLeaks: Membership Inference Attacks Against Text-to-Video Models}


\author{
Li Wang\textsuperscript{1,3,4}, 
Wenyu Chen\textsuperscript{1}, 
Ning Yu\textsuperscript{2}, 
Zheng Li\textsuperscript{1,3,4}\thanks{Corresponding authors}, 
Shanqing Guo\textsuperscript{1,3,4}\footnotemark[1]
\\[1ex]
\textsuperscript{1}\textit{School of Cyber Science and Technology, Shandong University;}
\textsuperscript{2}\textit{Eyeline Labs} \\
\textsuperscript{3}\textit{State Key Laboratory of Cryptography and Digital Economy Security, Shandong University} \\
\textsuperscript{4}\textit{Shandong Key Laboratory of Artificial Intelligence Security, Shandong University}
}

\maketitle

\begin{abstract}
 
The proliferation of powerful Text-to-Video (T2V) models, trained on massive web-scale datasets, raises urgent concerns about copyright and privacy violations. 
Membership inference attacks (MIAs) provide a principled tool for auditing such risks, yet existing techniques—designed for static data like images or text—fail to capture the spatio-temporal complexities of video generation. In particular, they overlook the sparsity of memorization signals in keyframes and the instability introduced by stochastic temporal dynamics.

In this paper, we conduct the first systematic study of MIAs against T2V models and introduce a novel framework \ourframework{}, which probes sparse-temporal memorization through two complementary signals: 
1) Spatial Reconstruction Fidelity (SRF), using a Top-K similarity to amplify spatial memorization signals from sparsely memorized keyframes, and 2) Temporal Generative Stability (TGS), which measures semantic consistency across multiple queries to capture temporal leakage. We evaluate \ourframework{} under three progressively restrictive black-box settings—supervised, reference-based, and query-only. Experiments on three representative T2V models reveal severe vulnerabilities: \ourframework{} achieves AUC of 82.92\% on AnimateDiff and 97.01\% on InstructVideo even in the strict query-only setting, posing a realistic and exploitable privacy risk. 
Our work provides the first concrete evidence that T2V models leak substantial membership information through both sparse and temporal memorization, establishing a foundation for auditing video generation systems and motivating the development of new defenses. Code is available at: \url{https://zenodo.org/records/17972831.} 
\end{abstract}

\section{Introduction}

\begin{figure}[!t]
\centering
\includegraphics[width=0.87\columnwidth]{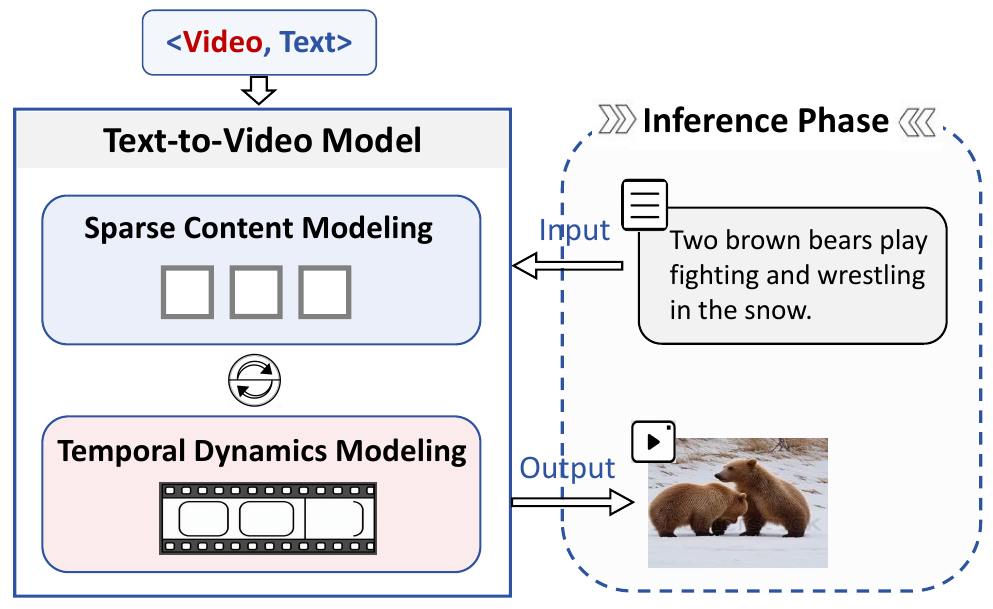} %
\caption{Illustration of the T2V generation process.} 
\label{fig:T2V_model}
\end{figure}

The advent of powerful Text-to-Video models (T2V), such as Sora~\cite{sora2024}, Kling~\cite{kling2024}, Luma~\cite{luma2024dream}, and Gen-3~\cite{runway2024gen3}, marks a new frontier in generative AI, enabling the creation of high-fidelity, dynamic video content directly from text prompts~\cite{sun2025t2v, wang2025lavie, blattmann2023stable, yang2024cogvideox, fan2025vchitect, wu2024boosting, guo2023animatediff}. However, this capability critically depends on training with massive datasets, often containing billions of internet-scraped videos that inevitably include private and unauthorized data~\cite{chen2024panda, bain2021frozen, wang2024vidprom, ju2024miradata}. 
The use of such data has already sparked significant controversy~\cite{miao2024t2vsafetybench}. 
For instance, in 2024, \textit{YouTube}'s CEO publicly stated that training OpenAI's \textit{Sora} with YouTube videos would ``clearly violate'' the platform's terms of service, while OpenAI's CTO declined to confirm whether such data had been used (reported by \textit{Bloomberg}\footnote{\url{https://www.bloomberg.com/news/articles/2024-04-04/youtube-says-openai-training-sora-with-its-videos-would-break-the-rules}}). 
The controversy has also spread internationally — after OpenAI launched Sora in the UK, debates intensified over AI training on copyrighted works and artist rights (reported by \textit{The Guardian}\footnote{\url{https://www.theguardian.com/technology/2025/feb/28/openai-sora-video-generation-uk-amid-copyright-row?utm_source=chatgpt.com}}).
This tension is mirrored in the open-source community, where a large video dataset curated for the Open-Sora project was removed from Hugging Face following a DMCA takedown request by the stock footage platform Pexels\footnote{\url{https://huggingface.co/datasets/hpcai-tech/open-sora-pexels-45k}}. 
Therefore, significant concerns have been raised regarding such data risks, motivating us to address a fundamental question, ``Is a given video used in training a T2V model?''


In this paper, we take the first step toward studying data risks of T2V models through membership inference attacks. 
The goal is to infer whether a given video is part of a T2V model's training dataset.
While MIAs have been widely explored for classifiers~\cite{li2021membership, lienhanced} and more recently for large language models (LLMs)~\cite{he2025towards} and text-to-image (T2I) models~\cite{hu2025membership}, extending them to T2V models introduces fundamentally new challenges. 
As illustrated in \autoref{fig:T2V_model}, T2V models are trained on video–text pairs to generate temporally coherent video sequences~\cite{menapace2024snap, feng2024tc, bar2024lumiere}. 
To this end, they adopt specialized mechanisms such as spatial-temporal sparse attention~\cite{fan2025vchitect,zhou2022magicvideo} to compress redundant frames, preserve spatial fidelity, and stabilize temporal dynamics. 
While these designs improve generation quality, they also complicate membership inference, creating unique attack surfaces beyond those in text- or image-based MIAs:

\begin{itemize}
\item \textbf{Challenge \filledcircled[\small]{1}: Sparsity of Content Memorization.} Video data is highly redundant, with many visually similar frames. 
To train efficiently, T2V models selectively memorize only sparse, informative anchors (e.g., keyframes~\cite{yuan2024instructvideo,huang2024vbench++}). This sparsity weakens membership signals, as the few memorized anchors are easily overwhelmed by the noise from generalized, non-memorized frames. Consequently, naive frame-wise similarity fails because it averages over all frames, drowning out the sparse memorization signal (see \autoref{sec:similarity metrics}).

\item \textbf{Challenge \filledcircled[\small]{2}: Dynamics of Temporal Memorization.} 
Beyond static appearance, videos encode motion over time. T2V models thus learn temporal dynamics in addition to spatial content~\cite{menapace2024snap, liao2024evaluation}. However, they generate stochastic motion textures rather than fixed trajectories. Natural motion variation introduces substantial pixel-level noise that overwhelms the subtle temporal cues distinguishing members from non-members. As a result, pixel-level tools such as optical flow~\cite{teed2020raft} fail to isolate reliable membership signals under this variability (see \autoref{ablation: motion}).


\end{itemize}

To address these challenges, we propose a novel sparse-temporal MIA framework, \ourframework{}, tailored to T2V models, which focuses on: (1) the fidelity of reconstructing sparsely memorized key content and (2) the stability of its temporal dynamics. 
Concretely, we design two complementary signals:
\begin{itemize}
\item \textbf{Signal \filledcircled[\small]{1}: Sparse Reconstruction Fidelity (SRF).} To overcome sparsity, SRF evaluates whether the model memorizes key content by introducing a \textit{Top-K Reconstruction Fidelity} metric. This metric compares generated frames against the most relevant keyframes of the target video, effectively acting as a matched filter that amplifies weak memorization signals otherwise obscured by frame redundancy.

\item \textbf{Signal \filledcircled[\small]{2}: Temporal Generative Stability (TGS).} To address temporal dynamics, TGS measures the stability of scene-level semantics across repeated generations via a \textit{Multi-Q Generative Stability} metric. This metric captures semantic consistency across generated frames over multiple queries, providing a reliable probe into whether the model has memorized temporal patterns of a video.

\end{itemize}

\begin{table}[t]
\centering
\caption{Adversary's knowledge under three threat models.}
\label{tab:threat_models}
\resizebox{0.87\linewidth}{!}{
\begin{tabular}{@{}l|cc|cc@{}}
\toprule[1.2pt] 
\multirow{2}{*}{\textbf{Threat Model}} & \multicolumn{2}{c|}{\textbf{Member Data}} & \multicolumn{2}{c}{\textbf{Non-Member Data}} \\
\cmidrule(lr){2-3} \cmidrule(lr){4-5}
 & Video & Text  & Video & Text  \\
\midrule
\textbf{Supervised} & \textcolor{green!70!black}{\Large\bfseries\checkmark} & \textcolor{red}{\Large\bfseries$\times$} & \textcolor{green!70!black}{\Large\bfseries\checkmark} & \textcolor{red}{\Large\bfseries$\times$} \\
\textbf{Reference-based} & \textcolor{red}{\Large\bfseries$\times$} & \textcolor{red}{\Large\bfseries$\times$} & \textcolor{green!70!black}{\Large\bfseries\checkmark} & \textcolor{red}{\Large\bfseries$\times$} \\
\textbf{Query-only} & \textcolor{red}{\Large\bfseries$\times$} & \textcolor{red}{\Large\bfseries$\times$} & \textcolor{red}{\Large\bfseries$\times$} & \textcolor{red}{\Large\bfseries$\times$} \\
\bottomrule[1.2pt]
\end{tabular}}
\end{table}

Building on SRF and TGS, we formalize our \ourframework{} and systematically evaluate it under three progressively restrictive black-box threat models (see \autoref{tab:threat_models}). 
Given a target video, an adversary first leverages a public video captioning model to obtain a surrogate text, which is then used to query the target T2V model. 
The generated videos are analyzed with SRF and TGS, and the results are fed into an inference module to determine membership.
Depending on the adversary’s capability, the module is instantiated in three scenarios (formally defined in \autoref{sec:threat_model}): (1) a trained classifier in the \textbf{supervised} setting, (2) a statistical anomaly detector in the \textbf{reference-based} setting, or (3) an unsupervised fusion in the \textbf{query-only} setting.

We conduct extensive experiments on three representative T2V models: AnimateDiff~\cite{guo2023animatediff}, Mira~\cite{mira2024github}, and InstructVideo~\cite{yuan2024instructvideo}, covering diverse architectural paradigms. 
Empirical evaluations show that \ourframework{} achieves strong performance across all settings, achieving AUC of 82.92\% on AnimateDiff and 97.01\% on InstructVideo, even under the most restrictive query-only scenario. 
This demonstrates that T2V models inevitably leak membership information through both sparse reconstruction fidelity and temporal generative stability.
In summary, our main contributions are as follows:
\begin{itemize}
\item 
We conduct the first comprehensive investigation of membership inference attacks against text-to-video models. We identify and formalize two domain-specific challenges: the sparsity of content memorization and the dynamics of temporal memorization.

\item 
We propose a novel MIA framework, \ourframework{}, introducing two complementary signals—Sparse Reconstruction Fidelity (SRF) and Temporal Generative Stability (TGS). These signals expose membership leakage by targeting keyframe fidelity and temporal stability, which conventional holistic or motion-based methods fail to capture.

\item 
We design an attack pipeline applicable under progressively restrictive black-box threat models, starting only from a target video without its ground-truth caption. Extensive experiments on three representative T2V models reveal severe vulnerabilities: \ourframework{} achieves strong performance across all scenarios, with AUC of 82.92\% on AnimateDiff and 97.01\% on InstructVideo, even in the most restrictive query-only setting. 
\end{itemize}

\section{Background \& Related Work}
\label{sec: background}
\subsection{Text-to-Video Generation Models}
\label{sec:t2v_background}

Text-to-Video (T2V) generation aims to synthesize temporally coherent videos from textual prompts~\cite{opensora2024,wu2024boosting,yuan2024instructvideo,sun2025t2v,blattmann2023align}. Recent breakthroughs are largely driven by diffusion models~\cite{ho2020denoising,song2020score}, particularly in latent space~\cite{rombach2022high}, where a denoiser is trained to iteratively recover clean video representations from noisy inputs conditioned on text~\cite{guo2023animatediff,luo2023videofusion}.  

Extending diffusion from static images to the spatio-temporal domain of video  has given rise to three representative paradigms~\cite{sun2024sora,zhang2025show,feng2024tc}: 
\textit{1) T2I adaptation with motion modules.} This paradigm builds upon powerful text-to-image backbones by freezing spatial layers and inserting lightweight temporal modules (e.g., temporal attention) for motion modeling~\cite{wu2023tune}. AnimateDiff~\cite{guo2023animatediff} is a canonical example, explicitly decoupling spatial appearance from temporal motion modeling.  
\textit{2) End-to-end spatio-temporal training.} Instead of reusing T2I backbones, these models train large spatio-temporal transformers directly on web-scale video–text datasets~\cite{bain2021frozen,wang2024vidprom,ju2024miradata}. Systems such as Mira~\cite{mira2024github}, Open-Sora~\cite{opensora2024}, and CogVideoX~\cite{yang2024cogvideox} belong to this category. They often employ factorized spatio-temporal attention to jointly capture visual and motion representations. 
\textit{3) Reward- or instruction-based fine-tuning.}  Inspired by alignment techniques in LLMs~\cite{wu2024boosting}, this line of work fine-tunes pre-trained T2V models with reward signals or human feedback. InstructVideo~\cite{yuan2024instructvideo} is a notable example, using image-based reward models to enhance visual appeal and text alignment.  

These paradigms—adaptation, end-to-end training, and fine-tuning—capture the major strategies in today’s T2V landscape. For our study, we select AnimateDiff, Mira, and InstructVideo as representative and auditable cases.  Unlike recent closed-source systems such as CogVideoX~\cite{yang2024cogvideox}, Hunyuan-Video~\cite{kong2024hunyuanvideo}, and Wan~\cite{wan2025wan}, whose training datasets are proprietary or undisclosed, our chosen models provide transparent data provenance, which is essential for verifiable membership inference studies. Importantly, despite their architectural differences, all share the common goal of modeling both static spatial fidelity and dynamic temporal evolution~\cite{menapace2024snap,liao2024evaluation,wu2023discovqa}, creating the attack surfaces we exploit.

\subsection{Membership Inference Attacks}
\label{sec:mia_evolution}

Membership Inference Attacks (MIAs) aim to determine whether a given sample was part of a model's training set~\cite{shokri2017membership}. 
Early studies focused on classification models, exploiting the observation that models typically yield higher confidence (or lower loss) on member samples they were trained on~\cite{li2021membership,lienhanced}.
The scope later has broadened to generative models, such as GANs~\cite{goodfellow2014generative,arjovsky2017wasserstein} and diffusion models~\cite{truong2025attacks}. 
For GANs, attacks exploited signals from the discriminator's output or the generator's ability to reconstruct a sample, as explored in GAN-Leaks~\cite{chen2020gan}. 
For diffusion models, particularly in the T2I domain, researchers have focused on the reconstruction error between queried images and their generated counterparts~\cite {wu2022membership}. 
These works established that reconstruction fidelity is a viable, albeit sometimes noisy, signal for membership.

With the rise of large-scale models, MIAs have faced new challenges. 
For LLMs~\cite{chang2024survey,kasneci2023chatgpt} and VLMs~\cite{tang2025video,momeni2023verbs}, which are often accessed only via black-box APIs, traditional loss-based signals are unavailable. 
This has spurred the development of ``label-only'' attacks that rely on model outputs alone~\cite{li2024membership, maini2024llm}. 
These attacks often probe for subtle behavioral differences, such as a model's consistency, robustness, or sensitivity to specific parameters~\cite{he2025towards, hu2025membership}. 
This evolution towards analyzing subtle behavioral artifacts inspires our design for T2V models.  

Despite these advances, the vulnerability of modern T2V models to MIAs has not been systematically studied. 
Prior works against T2I models are insufficient as they neglect the temporal dimension~\cite{menapace2024snap,liao2024evaluation}, while techniques for other modalities fail to capture the spatio-temporal generative process of video. 
To our knowledge, this work is the first to bridge this gap by systematically studying MIAs against T2V systems across diverse architectural paradigms. 

\section{Key Insights \& Signal Design}
\label{sec:insights}

To design an effective MIA on T2V models, we first revisit how these models memorize training data and identify signals that can expose such memorization. 
As discussed in \autoref{sec:t2v_background}, T2V models jointly optimize two objectives: (1) generating high-fidelity visual content within frames and (2) maintaining coherent motion across frames.
These dual objectives imply two complementary perspectives of memorization: (1) fidelity of visual details in key frames and (2) stability of temporal evolution. We now develop concrete signals that capture both aspects.

\subsection{Sparse Memorization: Reconstruction Fidelity}
\label{sec:spatial_probe}

A direct approach to detecting memorization is to compare a target video with one generated from its text prompt. However, naive frame-wise similarity is ineffective (see \autoref{sec:similarity metrics}), since most frames contain redundant content that obscures weak memorization signals. T2V models are more likely to memorize distinctive \textit{key anchors}—salient keyframes or regions—rather than entire sequences. 

\begin{figure}[t]
  \centering
  \begin{subfigure}[t]{0.49\columnwidth}
    \centering
    \includegraphics[width=\linewidth]{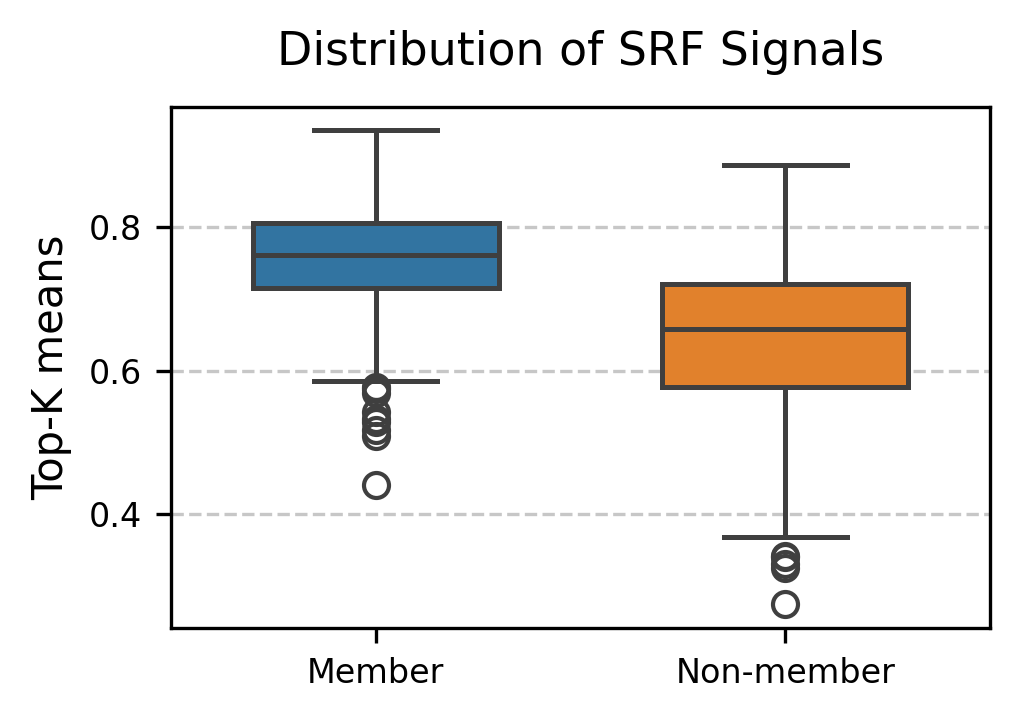}
    \subcaption{}
    \label{fig:core_signals_a}
  \end{subfigure}
  \hfill
  \begin{subfigure}[t]{0.49\columnwidth}
    \centering
    \includegraphics[width=\linewidth]{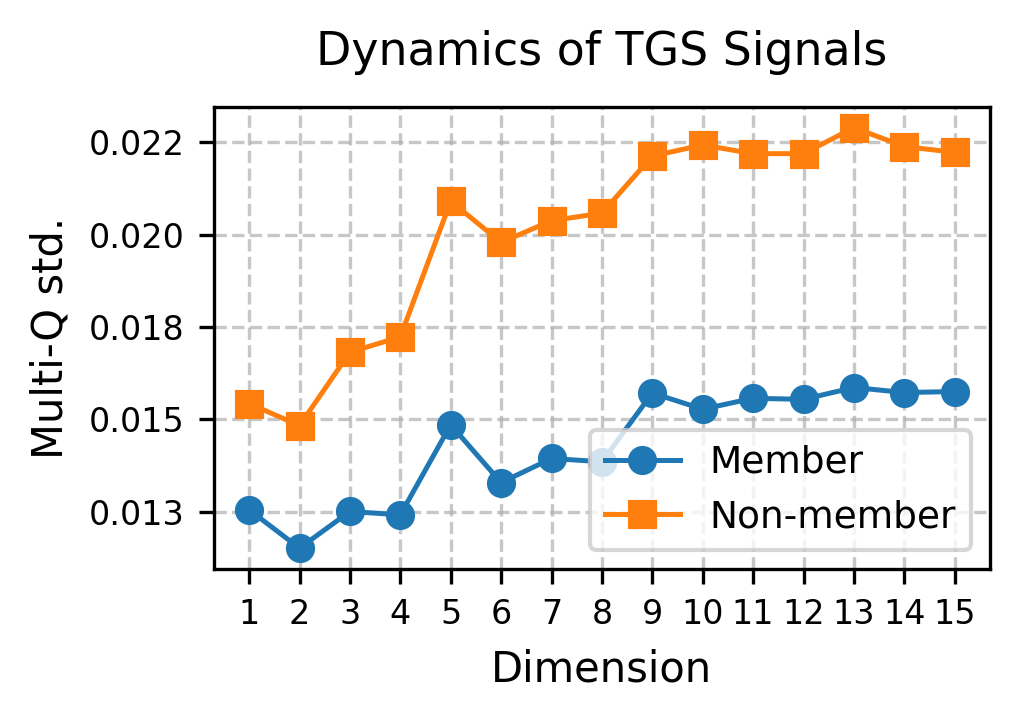}
    \subcaption{}
    \label{fig:core_signals_b}
  \end{subfigure}
  \caption{Differences between members and non-members under SRF and TGS signals. (a) Distribution of SRF scores computed from Top-$K$ similarities. (b) Per-dimension TGS instability across repeated generations.}
  \label{fig:core_signals}
\end{figure}

\paragraph{Signal \filledcircled[\small]{1}: Sparse Reconstruction Fidelity (SRF).} 
We design SRF to focus on these key anchors. We first extract keyframes from the target video using FFmpeg’s standard frame-selection tools~\cite{ffmpeg2000}, which provides a reproducible set of structural anchors in the video. For each generated frame, we compute its CLIP~\cite{radford2021learning} similarity with all extracted keyframes of the target video and average the Top-K scores. This acts as a matched filter that emphasizes memorized anchors while ignoring redundant frames. The overall SRF score is then averaged across generated frames. A higher SRF indicates stronger memorization of member videos. 
As shown in \autoref{fig:core_signals_a}, SRF produces a clear distributional shift between members and non-members, validating its effectiveness in probing sparse memorization.

Formally, let $v_{t}$ be a target video with $M$ keyframes $\mathcal{F}_{keys} = \{\mathbf{f}_1, \dots, \mathbf{f}_M\}$, and $\tilde{v}_{g}$ be the generated video with $N$ frames $\tilde{\mathcal{F}} = \{\tilde{\mathbf{f}}_1, \dots, \tilde{\mathbf{f}}_N\}$. For each generated frame $\tilde{\mathbf{f}}_i$, the SRF score is:
\begin{equation}
\label{eq:srf_per_frame}
SRF_{i} = \frac{1}{K} \sum_{j=1}^{K} \max_{k \in \{1..M\}}^{(j)} \cos(\tilde{\mathbf{f}}_i, \mathbf{f}_k)
\end{equation}
where $\max^{(j)}$ denotes the $j$-th largest value, and $\cos(\cdot, \cdot)$ denotes the cosine similarity.

The overall SRF score is $S_{SRF} = \frac{1}{N} \sum_{i=1}^N SRF_i$.  
In addition to this scalar, the full SRF vector $[SRF_1, \dots, SRF_N]$ can be used as a high-dimensional feature for supervised attacks.

\subsection{Temporal Memorization: Generative Stability}
\label{sec:temporal_probe}
The second perspective of memorization lies in motion patterns. However, measuring motion at the pixel level (e.g., optical flow) is unreliable (see \autoref{ablation: motion}), since T2V models learn stochastic motion textures rather than exact trajectories. Instead, the key lies in whether the model reproduces the same \textit{scene-level dynamics} consistently across generations.

\paragraph{Signal \filledcircled[\small]{2}: Temporal Generative Stability (TGS).}  
We propose TGS, which measures the stability of semantic scene evolution under repeated queries. For each generation, we first compute a frame-wise \textit{consistency vector} that captures background stability. By sampling the same text prompt $Q$ times, we then obtain $Q$ such vectors and measure per-dimension standard deviations. Member videos exhibit lower instability, reflecting stronger memorization.  
As shown in \autoref{fig:core_signals_b}, member videos exhibit consistently lower instability than non-members, establishing TGS as a robust probe of temporal memorization.

We follow the formulation of background consistency from~\cite{huang2024vbench++}. 
For a generated video with CLIP features $\{\tilde{\mathbf{f}}_0, \dots, \tilde{\mathbf{f}}_{N-1}\}$, the consistency score for frame $i>0$ is:  
\begin{equation}
\label{eq:tsc}
C_i = \tfrac{1}{2} \left( \cos(\tilde{\mathbf{f}}_i, \tilde{\mathbf{f}}_{i-1}) + \cos(\tilde{\mathbf{f}}_i, \tilde{\mathbf{f}}_0) \right)
\end{equation}
This yields a consistency vector $\mathcal{V} = [C_1, \dots, C_{N-1}]$. Across $Q$ generations, we compute the instability vector $\mathbf{s}_{instab} \in \mathbb{R}^{N-1}$, where each element is:
\begin{equation}
\label{eq:tgs_instab_vec}  
\mathbf{s}_{instab}[j] = \mathrm{StdDev}\big( \mathcal{V}_{1}[j], \dots, \mathcal{V}_{Q}[j] \big)
\end{equation}
The final TGS score is $S_{TGS} = \frac{1}{N-1} \sum_{j=1}^{N-1} \mathbf{s}_{instab}[j]$. Lower $S_{TGS}$ (higher stability) implies stronger memorization.

\begin{tcolorbox}[title=Remark,colback=gray!5!white,colframe=gray!75!black,fonttitle=\bfseries]
SRF captures whether the model memorizes salient visual anchors, while TGS measures whether it memorizes stable temporal evolution. 
Together, they provide complementary sparse-temporal signals that serve as the foundation of our attack.
\end{tcolorbox}

\begin{figure*}[t]
  \centering
  \includegraphics[width=0.92\linewidth]{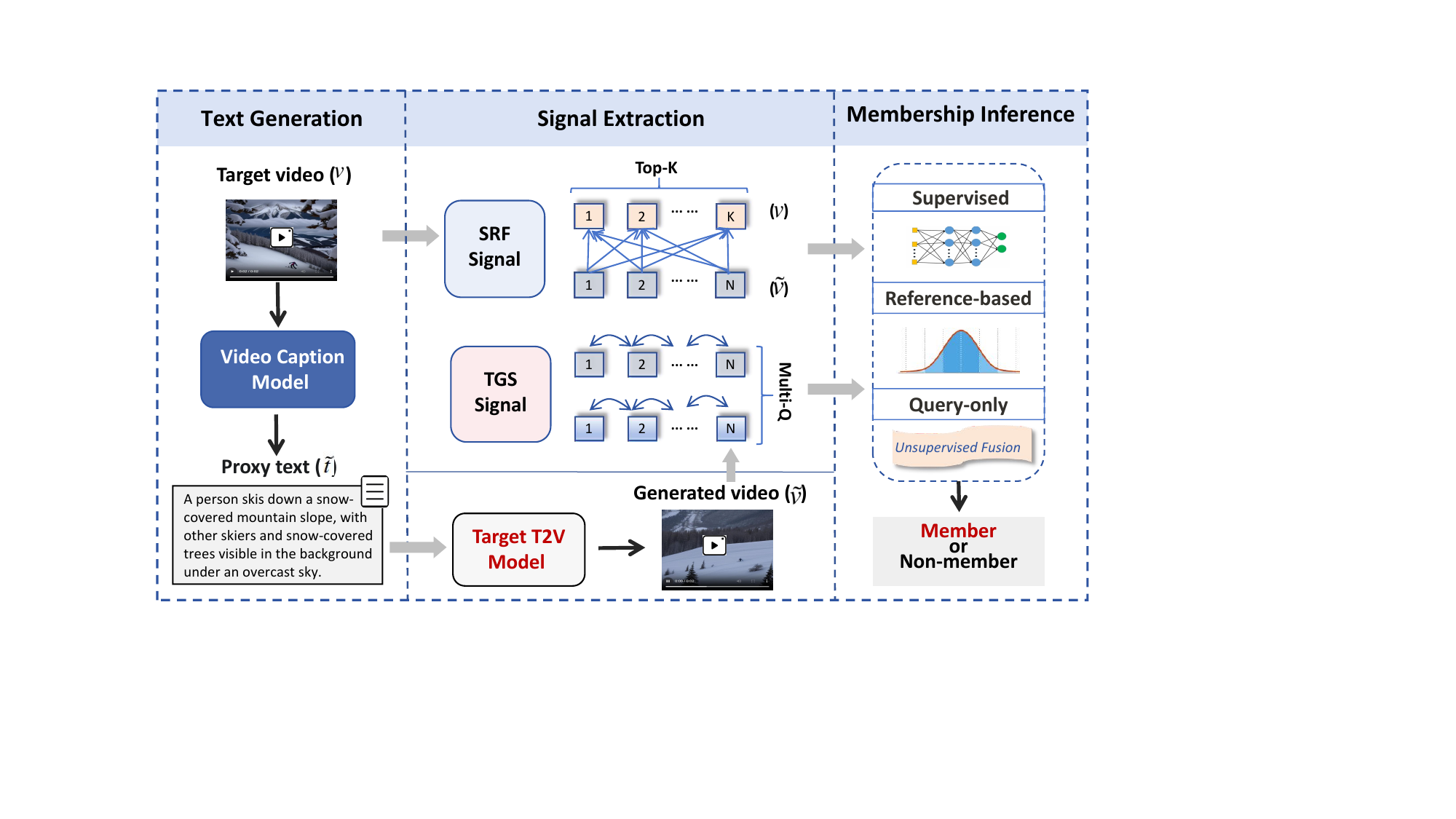}
 \caption{Overview of our sparse-temporal MIA framework. The attack begins with only a target video. A public video captioning tool generates a proxy text, which is then used to query the T2V model. The signal extraction stage computes SRF and TGS from the original and generated videos. Finally, the membership inference module, instantiated according to the threat model, outputs the final membership decision.}
  \label{fig:pipeline}
\end{figure*}

\section{The Sparse-Temporal MIA Framework}
\label{sec:framework}

Based on the insights from our initial explorations, we now formalize our sparse-temporal membership inference attack (MIA) framework against text-to-video (T2V) models. This section first defines the threat models and then presents the overall attack pipeline. 

\subsection{Threat Models}
\label{sec:threat_model}


We consider a black-box adversary who aims to determine whether a given \textbf{target video $v_{t}$} was included in the training set of a target T2V model $\mathcal{M}$. The adversary can query $\mathcal{M}$ with text prompts and observe the generated videos, but has no access to $\mathcal{M}$'s internal parameters, architecture, or training data. Critically, we assume the adversary does \emph{not} possess the ground-truth prompt paired with $v_{t}$ during training, as real-world auditing scenarios (e.g., creators verifying unauthorized use of their content) typically provide access only to the video itself rather than the original video–text pair. Following established MIA paradigms~\cite{hu2025membership}, we formalize three progressively restrictive threat models:

\begin{itemize}
    \item \textbf{Supervised Inference.} This setting establishes a theoretical upper bound on the attack performance. The adversary is assumed to have a \textit{shadow dataset} $\mathcal{D}_{shadow}$ containing samples explicitly labeled as members or non-members of the target model's training set.
    \item \textbf{Reference-based Inference.} In this more realistic model, the adversary no longer has access to labeled member samples but possesses a \textit{reference set} $\mathcal{D}_{ref}$ of confirmed non-member samples.
    \item \textbf{Query-only Inference.} This is the most restrictive and practical model. The adversary operates under a \textit{zero-knowledge} assumption, having neither a shadow nor a reference dataset. The inference must be made solely based on the query results of the target video itself.
\end{itemize}




\subsection{Attack Framework Overview}
\label{sec:attack_pipeline}

Our attack pipeline, illustrated in \autoref{fig:pipeline}, is a modular framework consisting of three main steps: Text Generation, Signal Extraction, and Membership Inference.

\begin{itemize}
    \item \textbf{Step \filledcircled[\small]{1} – Text Generation.} The attack realistically begins with only a target video $v$. The adversary first feeds $v$ into a publicly available video captioning model, $\mathcal{C}$, to generate a descriptive text, $\tilde{t} = \mathcal{C}(v)$. This proxy text serves as the input for querying the target T2V model. This step makes our attack highly practical as it removes the often unrealistic assumption of knowing the ground-truth prompt (see \autoref{sec:ablation_captioning} for the impact of caption source and quality).

    \item \textbf{Step \filledcircled[\small]{2} – Signal Extraction.} Using the generated prompt $\tilde{t}$, the adversary queries the target model to obtain a generation video $\tilde{v} = \mathcal{M}(\tilde{t})$. This stage then computes our two core signals based on the original video $v$ and the generated video $\tilde{v}$: the Sparse Reconstruction Fidelity (SRF) and the Temporal Generative Stability (TGS).

    \item \textbf{Step \filledcircled[\small]{3} – Membership Inference.} The extracted signals are then passed to a membership inference module, which outputs a final membership decision. The implementation of this module depends on the threat models. It can be instantiated as a trained classifier in the supervised scenario, a statistical anomaly scorer in the reference-based scenario, or an unsupervised fusion in the query-only scenario. The specific implementations for each instantiation are detailed in \autoref{sec:supervised_inference}, \autoref{sec:reference_inference}, and \autoref{sec:query_inference}. 
\end{itemize}

\section{Supervised Inference}
\label{sec:supervised_inference}

\subsection{Assumptions and Intuition}
\label{sec:supervised_intuition}
We begin with the \textbf{Supervised Inference} threat model, which provides a theoretical upper bound on potential information leakage. In this setting, the adversary is assumed to have a shadow dataset $\mathcal{D}_{\text{shadow}} = \{(v_i, \tilde{t}_i, y_i)\}_{i=1}^{N_s}$, where each sample is labeled as a member ($y_i=1$) or non-member ($y_i=0$) of the target model’s training set. The key intuition is that a supervised classifier, trained on this labeled data, can fully exploit the high-dimensional sparse-temporal signals to learn discriminative boundaries between members and non-members.

\subsection{Attack Implementation}
\label{sec:supervised_methodology}
The supervised attack leverages both SRF and TGS signals to train a classifier that can then infer the membership of target samples. The procedure consists of two stages: feature construction, and classifier training and inference (see Algorithm \ref{alg:supervised} in \autoref{app:algorithms}).

\paragraph{Feature Construction.}
To capture the rich information, we represent each sample by its full underlying signal vectors rather than scalar scores ($S_{SRF}$, $S_{TGS}$) defined in \autoref{sec:insights}. For each sample $(v_i, \tilde{t}_i)$ in the shadow dataset, we extract:
\begin{itemize}
    \item \textbf{SRF vector:} $\mathbf{v}_{srf} = [SRF_1, \dots, SRF_N] \in \mathbb{R}^N$, where each element is the Top-K fidelity of a generated frame (\autoref{eq:srf_per_frame}).
    \item \textbf{TGS vector:} $\mathbf{s}_{instab} \in \mathbb{R}^{N-1}$, where each element is the standard deviation of temporal consistency score at frame index $j$ across $Q$ generations (\autoref{eq:tgs_instab_vec}).
\end{itemize}

These vectors are then concatenated to form a single, high-dimensional feature vector $\mathbf{x} \in \mathbb{R}^{2N-1}$ for each sample. This representation preserves both spatial and temporal patterns essential for distinguishing members from non-members.

\begin{figure*}[t!]
  \centering
  \begin{subfigure}[t]{0.31\textwidth}
    \centering
    \includegraphics[width=\linewidth]{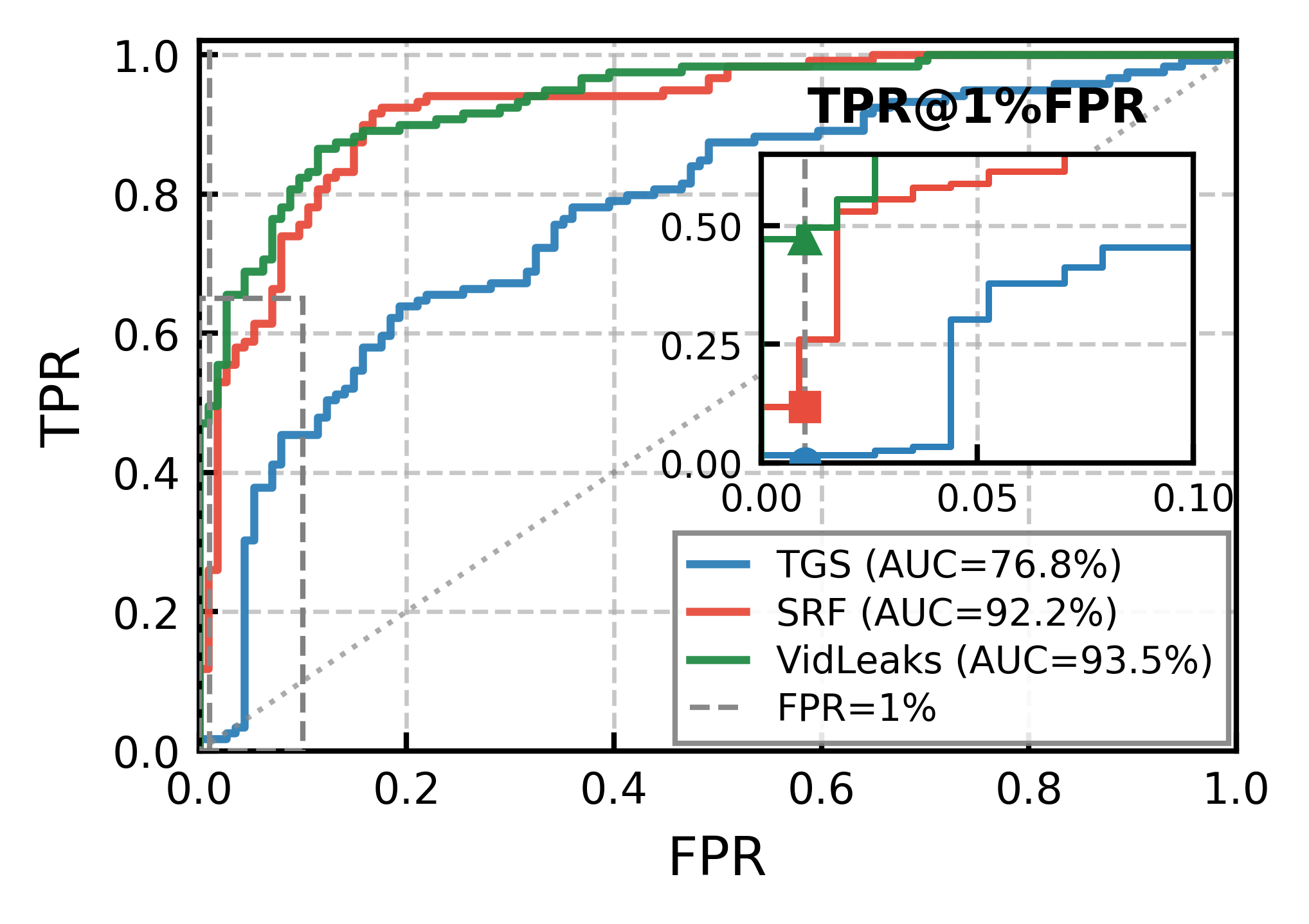}
    \subcaption{AnimateDiff}
    \label{fig:attack1_animatediff_roc}
  \end{subfigure}
  \hfill
  \begin{subfigure}[t]{0.31\textwidth}
    \centering
    \includegraphics[width=\linewidth]{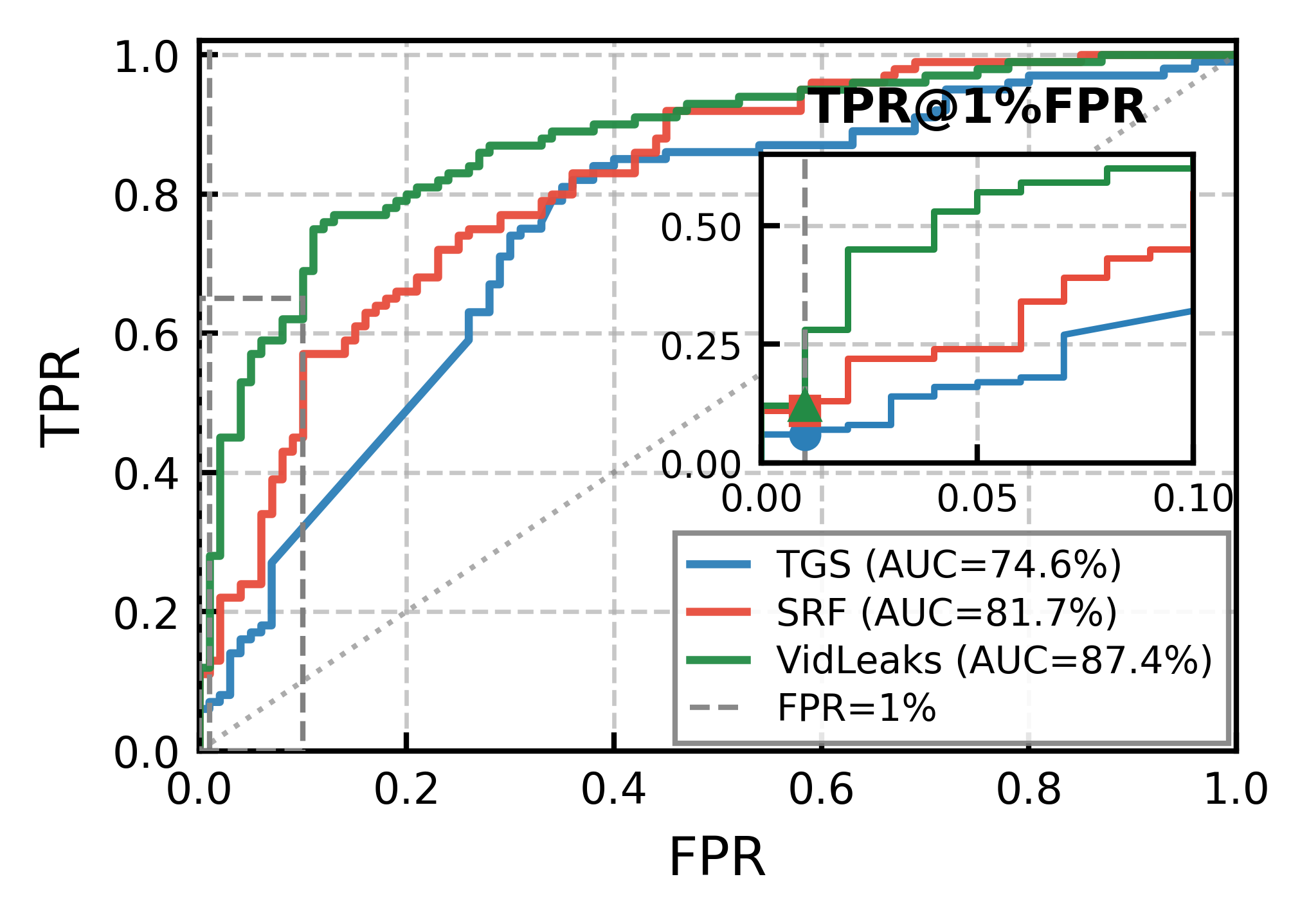}
    \subcaption{Mira}
    \label{fig:attack1_mira_roc}
  \end{subfigure}
  \hfill
  \begin{subfigure}[t]{0.31\textwidth}
    \centering
    \includegraphics[width=\linewidth]{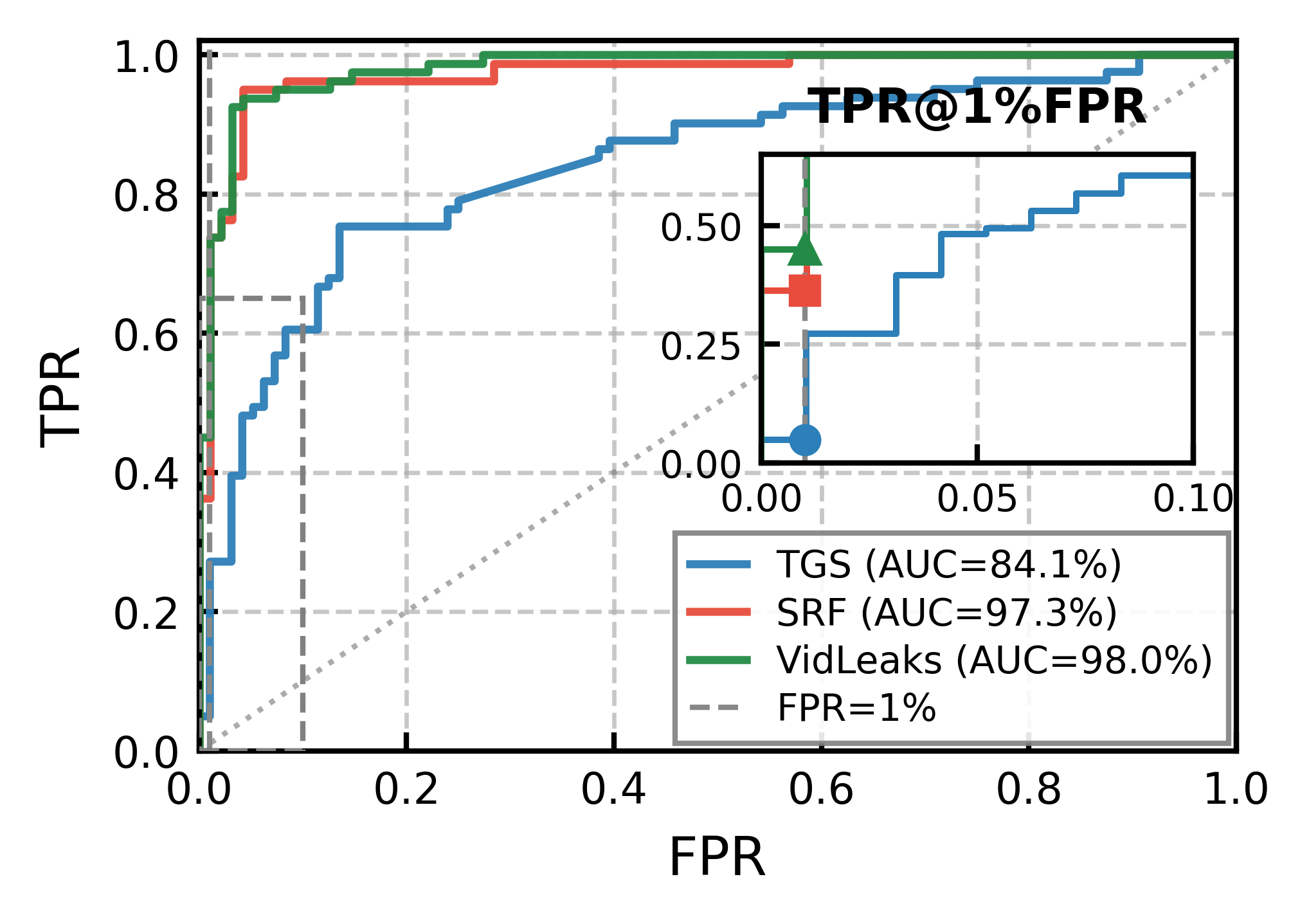}
    \subcaption{InstructVideo}
    \label{fig:attack1_instrcut_roc}  
  \end{subfigure}
  \caption{ROC curves for the supervised attack on different T2V models.}
  \label{fig:supervised_roc}
\end{figure*}

\paragraph{Classifier Training and Inference.} 
Using the labeled feature vectors $\{\mathbf{x}_i, y_i\}_{i=1}^{N_s}$, the adversary trains a Multi-Layer Perceptron (MLP) to serve as the attack model, $\mathcal{A}_\theta$. The MLP is trained to minimize the loss between its predictions and the ground-truth labels, thereby learning the discriminative patterns between member and non-member samples. 
Once trained, $\mathcal{A}_\theta$ can process the feature vector $\mathbf{x}$ of any target video and outputs a membership probability $p = \mathcal{A}_\theta(\mathbf{x}) \in [0,1]$, which is used to decide whether the video belongs to the model’s training set. 



\subsection{Experimental Setting}

\paragraph{Target T2V Models.}
To ensure coverage of the main T2V paradigms introduced in \autoref{sec:t2v_background}, we evaluate three representative open-source models:
\begin{itemize}
    \item \textbf{AnimateDiff}~\cite{guo2023animatediff}: a canonical T2I-adaptation model that freezes image backbones and inserts lightweight motion modules.
    \item \textbf{Mira}~\cite{mira2024github}: an end-to-end spatio-temporal transformer trained on the curated MiraData dataset. Mira (Mini-Sora) represents an initial foray into high-quality, long-duration video generation in the style of Sora.
    \item \textbf{InstructVideo}~\cite{yuan2024instructvideo}: a reward-fine-tuned model aligned with human preference signals.
\end{itemize}
These models jointly cover the adaptation, end-to-end, and alignment paradigms of current T2V systems. All three models have publicly documented training corpora, allowing us to reliably determine membership for rigorous MIA evaluation. 

\paragraph{Datasets.}
Member samples are drawn directly from the publicly documented training sources of each target model: WebVid-10M~\cite{bain2021frozen} for AnimateDiff and InstructVideo, and MiraData~\cite{ju2024miradata} for Mira, ensuring unambiguous membership labels. Non-member samples are taken from Panda-70M~\cite{chen2024panda}, a high-quality dataset disjoint from all training sources, and are randomly sampled for evaluation.    
In the supervised scenario, the shadow dataset for each T2V model contains approximately 500 member and 500 non-member videos. All datasets are stratified and split 8:2 into training and testing sets.

\paragraph{Video Caption Model.}
Although ground-truth text prompts are available in the training datasets, we adopt a more realistic setting where target videos may not always come with captions. To query target models, we therefore employ a proxy captioner (Gemini Pro via Google AI Studio~\cite{GoogleAIStudio}) to generate descriptive prompts. Crucially, \ourframework{} maintains strong performance across diverse caption sources and qualities (see \autoref{sec:ablation_captioning}), ensuring our pipeline's practicality when ground-truth or high-quality captions are unavailable. 


\paragraph{Attack Model and Training.}
For the supervised setting, the attack model is a MLP as defined in \autoref{sec:supervised_methodology}, with two hidden layers using ReLU activations and dropout. We train the model with binary cross-entropy loss and early stopping based on validation AUC.  

\paragraph{Evaluation Metrics.}
We evaluate attack performance using three standard metrics:
\begin{itemize}
    \item \textbf{AUC (Area Under ROC Curve)}, a threshold-free measure of separability;
    \item \textbf{Balanced Accuracy}, which accounts for balanced performance across classes;
    \item \textbf{TPR@1\%FPR}, the true positive rate at a 1\% false positive rate.
\end{itemize}
We emphasize TPR@1\%FPR, following the paradigm established by Carlini et al.~\cite{carlini2022membership} and adopted in recent studies~\cite{he2025towards}, as it is particularly crucial for quantifying the practical risk of high-confidence membership inference attacks.

\subsection{Experimental Results}
\paragraph{Overall Performance.}
\autoref{tab:supervised_results} reports the results of our supervised attack across three representative T2V models. By jointly leveraging SRF and TGS vectors, our method achieves consistently strong performance, establishing a theoretical upper bound on membership leakage. On AnimateDiff, the attack reaches 93.46\% AUC and nearly 50\% TPR@1\%FPR, showing that almost half of the member videos can be identified with extremely high confidence at only 1\% false positives. Even on Mira, which is trained end-to-end and thus considered harder to attack, our method achieves 87.45\% AUC, significantly above chance. InstructVideo appears especially vulnerable, with 98.04\% AUC and 45\% TPR@1\%FPR, underscoring that preference-aligned fine-tuning can substantially amplify memorization risks. These results highlight the generality of our attack and its ability to extract sensitive membership information across diverse T2V architectures.

\begin{table}[t!]
\centering
\caption{Performance of the supervised attack across different T2V models.}
\label{tab:supervised_results}
\resizebox{\linewidth}{!}{
\begin{tabular}{l l ccc}
\toprule
\textbf{Target Model} & \textbf{Method} 
& \textbf{AUC(↑)} & \textbf{TPR@1\%FPR(↑)} & \textbf{ACC(↑)} \\
\midrule
\multirow{3}{*}{AnimateDiff} 
  & SRF  & 92.19\% & 26.89\% & 87.46\% \\
  & TGS  & 76.75\% & 2.52\% & 72.03\% \\
  & \cellcolor{gray!10}\ourframework{} & \cellcolor{gray!10}\textbf{93.46\%} & \cellcolor{gray!10}\textbf{49.58\%} & \cellcolor{gray!10}\textbf{87.58\%} \\
\midrule
\multirow{3}{*}{Mira}        
  & SRF  & 81.73\% & 11.00\% & 74.50\% \\
  & TGS  & 74.64\% & 6.00\% & 73.00\% \\
  & \cellcolor{gray!10}\ourframework{} & \cellcolor{gray!10}\textbf{87.45\%} & \cellcolor{gray!10}\textbf{12.00\%} & \cellcolor{gray!10}\textbf{82.00\%} \\
\midrule
\multirow{3}{*}{InstructVideo} 
  & SRF  & 97.29\% & 36.25\% & 94.44\% \\
  & TGS  & 84.07\% & 4.94\% & 80.88\% \\
  & \cellcolor{gray!10}\ourframework{} & \cellcolor{gray!10}\textbf{98.04\%} & \cellcolor{gray!10}\textbf{45.00\%} & \cellcolor{gray!10} \textbf{94.77\%} \\
\bottomrule
\end{tabular}}
\end{table}

\paragraph{Contribution of Each Signal.}
The ablation results in \autoref{tab:supervised_results} further validate our key insight: both SRF and TGS independently expose membership leakage. SRF achieves up to 97.29\% AUC (InstructVideo), reflecting the model’s tendency to memorize key visual anchors. TGS also provides non-trivial predictive power (e.g., 84.07\% AUC on InstructVideo), confirming that temporal stability encodes valuable membership cues despite being a noisier signal. Crucially, fusing SRF and TGS consistently improves performance across all models, boosting AUC by 1–6\% and substantially increasing TPR@1\%FPR. This demonstrates that the two signals capture complementary aspects of memorization, sparse spatial fidelity and temporal stability, that a supervised classifier can effectively combine to form a robust decision boundary. The ROC curves in \autoref{fig:supervised_roc} corroborate this, with the fused model’s curve dominating in the low-FPR region, where practical attacks are most relevant.



\section{Reference-based Inference}
\label{sec:reference_inference}

\begin{figure*}[t!]
  \centering
  \begin{subfigure}[t]{0.32\textwidth}
    \centering
    \includegraphics[width=\linewidth]{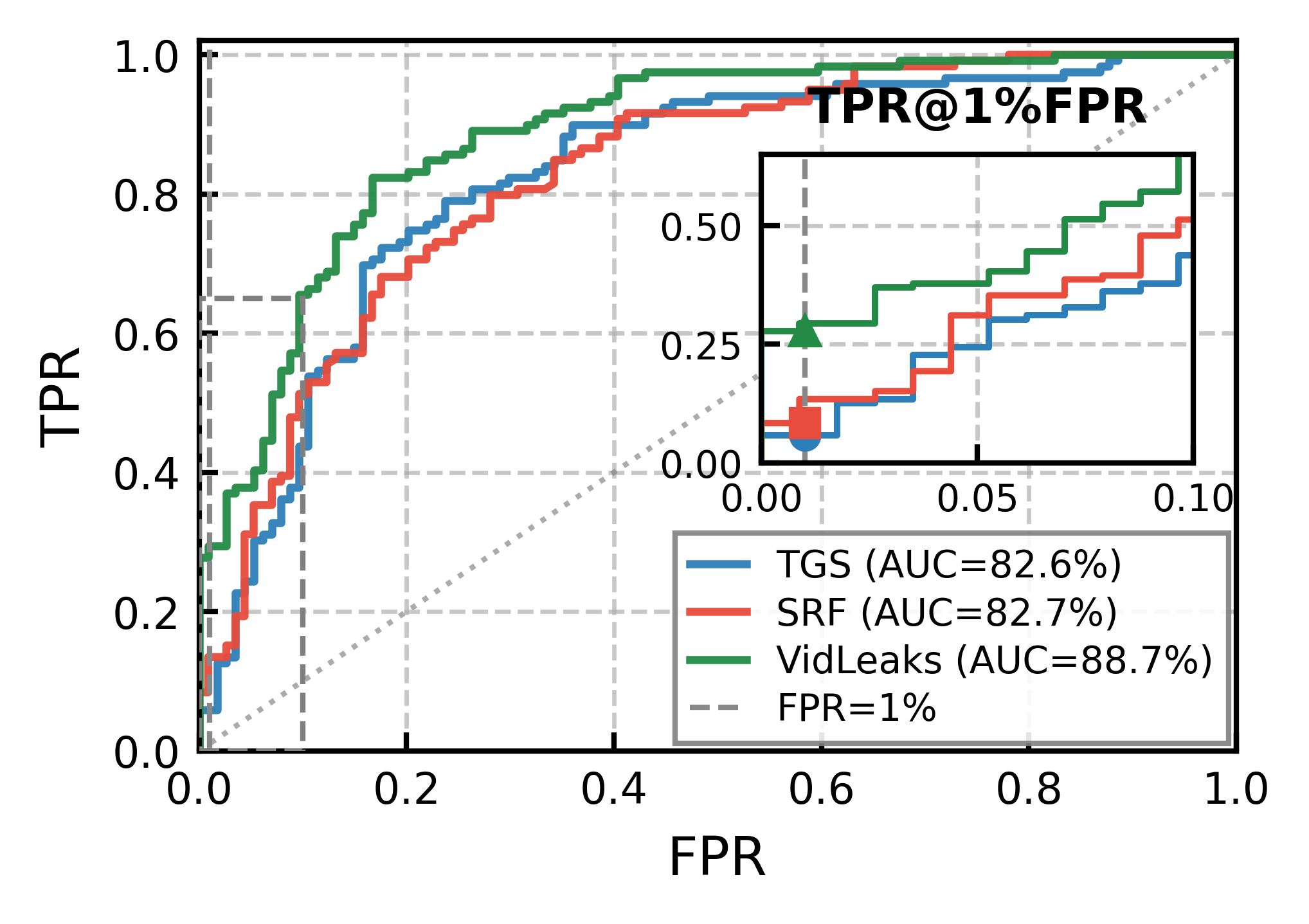}
    \subcaption{AnimateDiff}
    \label{fig:attack2_animatediff_roc}
  \end{subfigure}
  \hfill
  \begin{subfigure}[t]{0.32\textwidth}
    \centering
    \includegraphics[width=\linewidth]{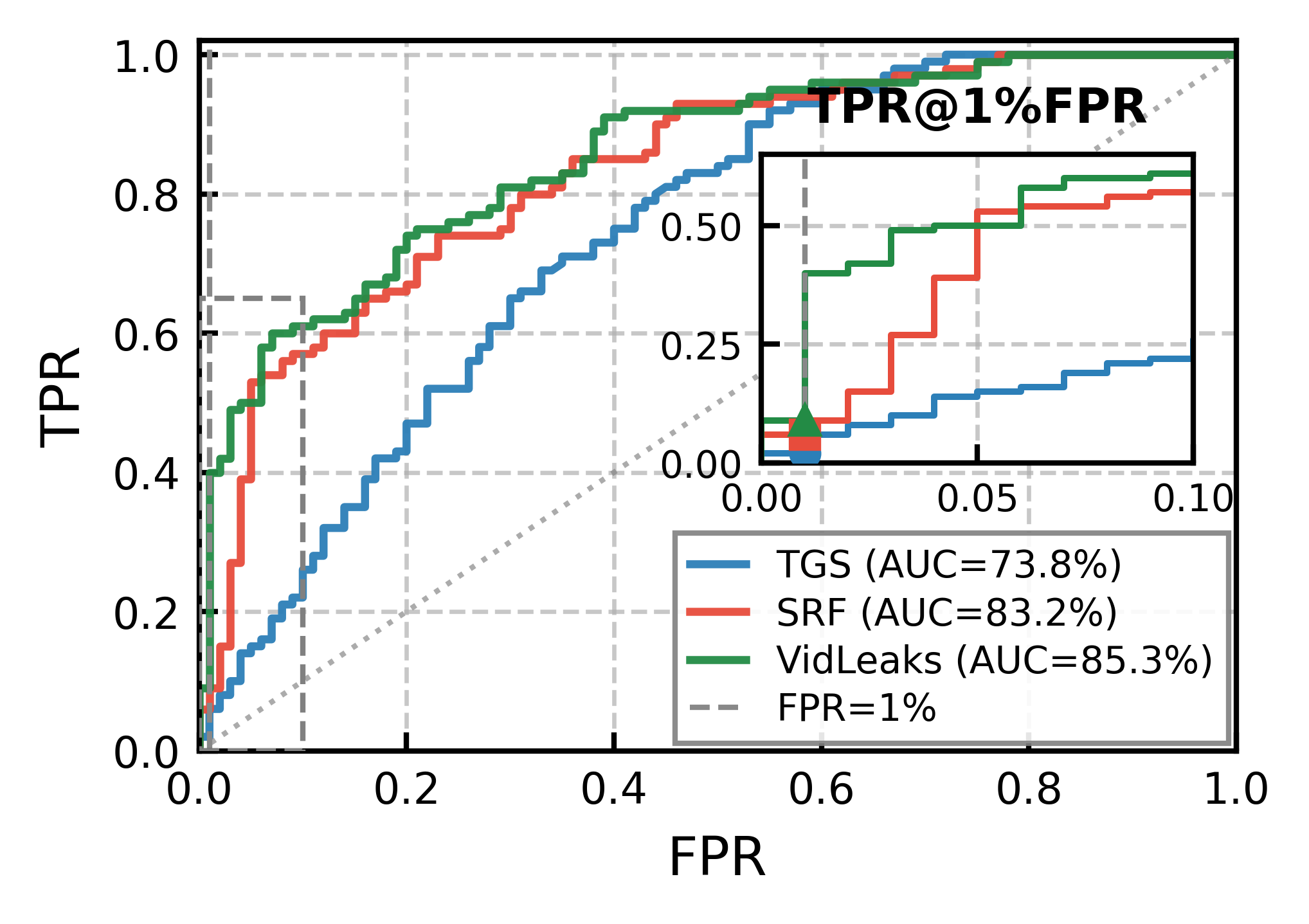}
    \subcaption{Mira}
    \label{fig:attack2_mira_roc}
  \end{subfigure}
  \hfill
  \begin{subfigure}[t]{0.32\textwidth}
    \centering
    \includegraphics[width=\linewidth]{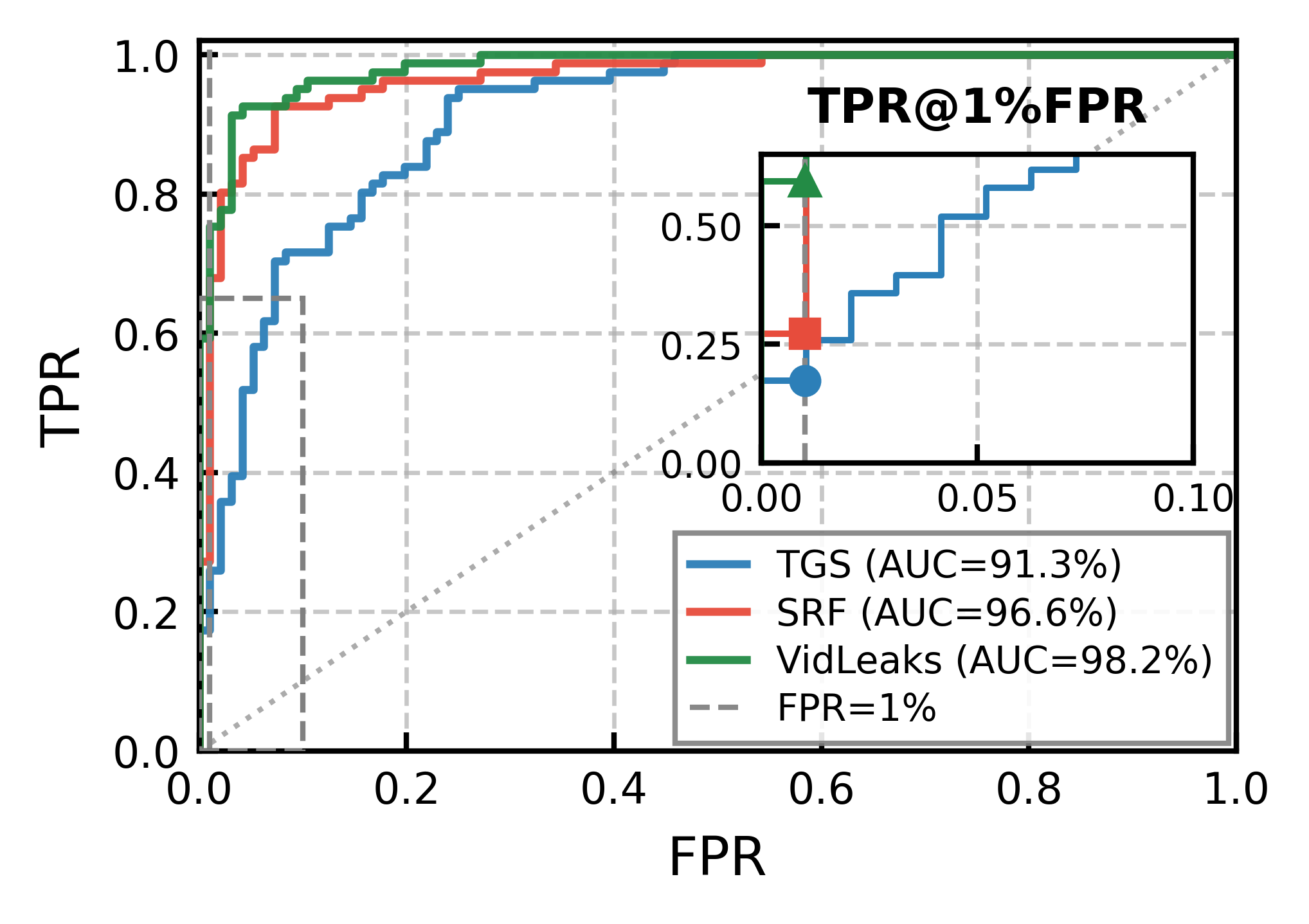}
    \subcaption{InstructVideo}
    \label{fig:attack2_instrcut_roc}  
  \end{subfigure}
  \caption{ROC curves for the reference-based attack on different T2V models.}
  \label{fig:reference_roc}
\end{figure*}

\subsection{Assumptions and Intuition}
\label{sec:reference_intuition}
We relax the strong assumption of a labeled shadow dataset and consider a more realistic threat model: \textbf{Reference-Based Inference}. This scenario assumes an adversary who, while lacking access to verified member samples, possesses a confirmed \textit{non-member reference set}, $\mathcal{D}_{\text{ref}} = \{(v_i, \tilde{t}_i)\}_{i=1}^{N_r}$. This setting is practical and realistic, since such non-member data can often be readily obtained (e.g., videos created after the target T2V model's knowledge cutoff date), whereas a labeled member set is typically inaccessible.

The key intuition of this attack is to use a reference set to establish a statistical baseline for non-member behavior. Membership is then inferred by measuring how much a target sample's signals deviate from this baseline. A significant deviation from the non-member distribution serves as strong evidence for membership. We formalize this as a statistical anomaly detection problem, where non-members form the distribution of normal data, and members are treated as out-of-distribution anomalies.

\subsection{Attack Implementation}
\label{sec:ref_methodology}
The reference-based attack replaces the supervised classifier with a statistical anomaly scoring mechanism. The procedure consists of three stages: calibration on the reference set, anomaly scoring for the target sample, and signal fusion and inference (see Algorithm \ref{alg:reference_based} in \autoref{app:algorithms}).  

\paragraph{Calibration on the Reference Set.}
The first step is to establish a statistical baseline for non-member behavior. The adversary computes the scalar SRF score and TGS score for all samples in the non-member reference set $\mathcal{D}_{\text{ref}}$. From these two sets of scores, they compute the mean ($\mu_{srf}, \mu_{tgs}$) and standard deviation ($\sigma_{srf}, \sigma_{tgs}$), characterizing the ``normal'' range for each signal.

\paragraph{Anomaly Scoring.}
For a target video, the adversary  extracts its scalar $S_{SRF}$ and $S_{TGS}$ scores. These raw scores are then transformed into normalized anomaly scores using the statistics derived from the reference set. We employ the Z-score for this normalization, which measures how many standard deviations a score is from the non-member mean.

For the $S_{SRF}$ score, where higher values indicate membership, the anomaly score $\mathcal{A}_{SRF}$ is its Z-score:
\begin{equation}
\label{eq:ascore_srf}
\mathcal{A}_{SRF} = \frac{S_{SRF} - \mu_{srf}}{\sigma_{srf}}
\end{equation}
Conversely, for the $S_{TGS}$ score, where lower values (higher stability) indicate membership, the anomaly score $\mathcal{A}_{TGS}$ is its negative Z-score to align the directionality:
\begin{equation}
\label{eq:ascore_tgs}
\mathcal{A}_{TGS} = - \frac{S_{TGS} - \mu_{tgs}}{\sigma_{tgs}}
\end{equation}
For both anomaly scores, a larger positive value signifies a greater deviation from non-member behavior and thus a higher likelihood of membership.

\paragraph{Signal Fusion and Inference.}
As SRF and TGS signals capture complementary aspects of memorization (\autoref{sec:insights}), we combine them into a single robust membership score via linear fusion:  
\begin{equation}
\label{eq:fusion_reference}
\mathcal{S}_{final} = w_{srf} \cdot \mathcal{A}_{SRF} + w_{tgs} \cdot \mathcal{A}_{TGS}
\end{equation}
where $w_{srf}$ and $w_{tgs}$ are weighting parameters. This final score $\mathcal{S}_{final}$ is then used to infer membership, typically by comparing it against a decision threshold.

\begin{table}[t!]
\centering
\caption{Performance of the reference-based attack across different T2V models.}
\label{tab:reference_results}
\resizebox{\linewidth}{!}{
\begin{tabular}{l l ccc}
\toprule
\textbf{Target Model} & \textbf{Method} 
& \textbf{AUC(↑)} & \textbf{TPR@1\%FPR(↑)} & \textbf{ACC(↑)} \\
\midrule
\multirow{3}{*}{AnimateDiff} 
  & SRF  & 82.70\% & 8.40\% & 75.88\% \\
  & TGS  & 82.63\% & 5.88\% & 77.65\% \\
  & \cellcolor{gray!10}\ourframework{} & \cellcolor{gray!10}\textbf{88.68\%} & \cellcolor{gray!10}\textbf{27.73\%} & \cellcolor{gray!10}\textbf{82.84\%} \\
\midrule
\multirow{3}{*}{Mira}        
  & SRF  & 83.15\% & 6.00\% & 75.50\% \\
  & TGS  & 73.75\% & 2.00\% & 68.50\% \\
  & \cellcolor{gray!10}\ourframework{} & \cellcolor{gray!10}\textbf{85.34\%} & \cellcolor{gray!10}\textbf{9.00\%} & \cellcolor{gray!10}\textbf{77.00\%} \\
\midrule
\multirow{3}{*}{InstructVideo} 
  & SRF  & 96.62\% & 27.16\% & 92.65\% \\
  & TGS  & 91.31\% & 17.28\% & 85.03\% \\
  & \cellcolor{gray!10}\ourframework{} & \cellcolor{gray!10}\textbf{98.17\%} & \cellcolor{gray!10}\textbf{59.26\%} & \cellcolor{gray!10}\textbf{94.21\%} \\
\bottomrule
\end{tabular}}
\end{table}

\subsection{Experimental Setting}
We evaluate the reference-based attack on the same target models, datasets, and proxy captioning setup as in \autoref{sec:supervised_inference}, and report results with the same metrics (AUC, Balanced Accuracy, and TPR@1\%FPR). The key difference lies in the attack configuration: instead of a labeled shadow dataset, the adversary is given only a disjoint non-member reference set. For each model, 80\% of the non-member pool is used to construct $\mathcal{D}_{ref}$ (for calibrating the SRF and TGS score distributions), while the remaining 20\% is reserved for evaluation alongside the member samples. For signal fusion (\autoref{eq:fusion_reference}), we adopt a conservative scheme with approximately equal weights ($w_{srf} \approx w_{tgs}$), avoiding hyperparameter tuning and providing a clean baseline for assessing the effectiveness of our fusion strategy. 

\subsection{Experimental Results}
\label{sec6: experiment}

\autoref{tab:reference_results} reports the results of the reference-based attack. Despite the absence of labeled member data, our fused method remains highly effective. On AnimateDiff, it achieves 88.68\% AUC and 27.73\% TPR@1\%FPR, showing that a non-member reference baseline alone suffices to reveal strong membership leakage. Similar patterns hold on Mira (85.34\% AUC) and InstructVideo (98.17\% AUC), with the latter reaching nearly 60\% TPR@1\%FPR. While overall performance is naturally below the supervised upper bound, these results confirm that significant privacy risks persist even under this more realistic threat model.  

Both SRF and TGS signals provide predictive power individually, with SRF generally performing better, but the fused approach consistently delivers the strongest results—raising AUC by up to 2–6\% and substantially improving TPR@1\%FPR. The ROC curves in \autoref{fig:reference_roc} further validate this trend, showing that even with only a non-member baseline, the fused score maintains a clear advantage over single-signal methods, particularly in the critical low-FPR region. This demonstrates that our \ourframework{}, which jointly exploits SRF and TGS, provides strong evidence of memorization even when inference relies solely on non-member statistics.

\begin{figure}[t!]
  \centering
  \includegraphics[width=0.97\linewidth]{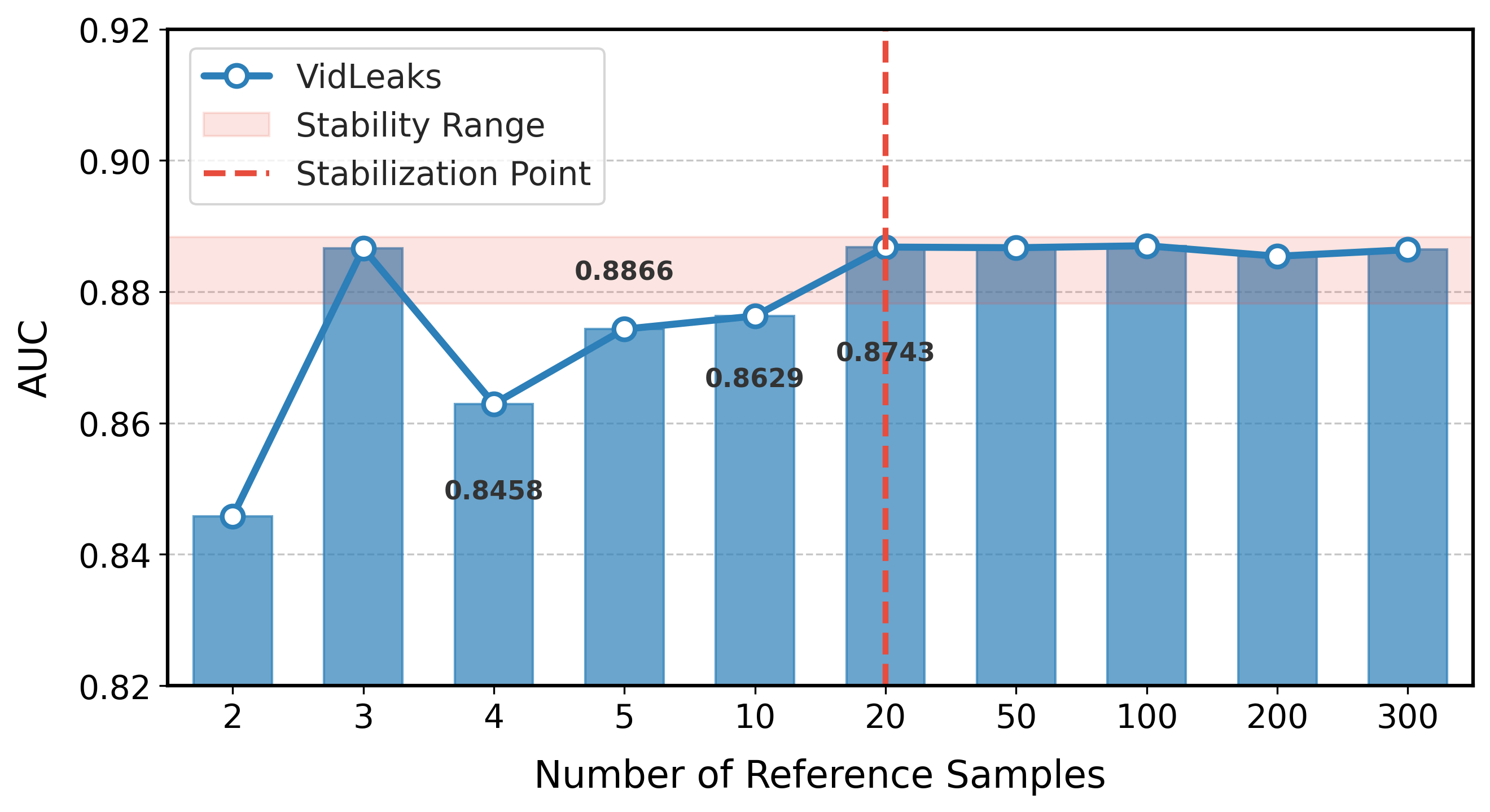}
\caption{Impact of reference set size on attack performance. }
\label{fig:ref_size_impact}
\end{figure}

\paragraph{Impact of Reference Set Size.}
We further examine how the size of the non-member reference set affects the attack performance. \autoref{fig:ref_size_impact} shows the AUC on AnimateDiff when varying the reference set from 2 to 300 samples. The attack already achieves an AUC of 88.66\% with only 3 samples, and performance quickly stabilizes once the reference set reaches around 20 samples, maintaining AUC above 88.5\% thereafter.  

These results demonstrate the data efficiency of our approach: only a handful of non-member videos are sufficient to construct a stable baseline for anomaly scoring. Consequently, the reference-based attack remains highly practical, as an adversary does not need access to a large dataset to reliably expose membership leakage.

\begin{figure*}[t!]
  \centering
  \begin{subfigure}[t]{0.32\textwidth}
    \centering
    \includegraphics[width=\linewidth]{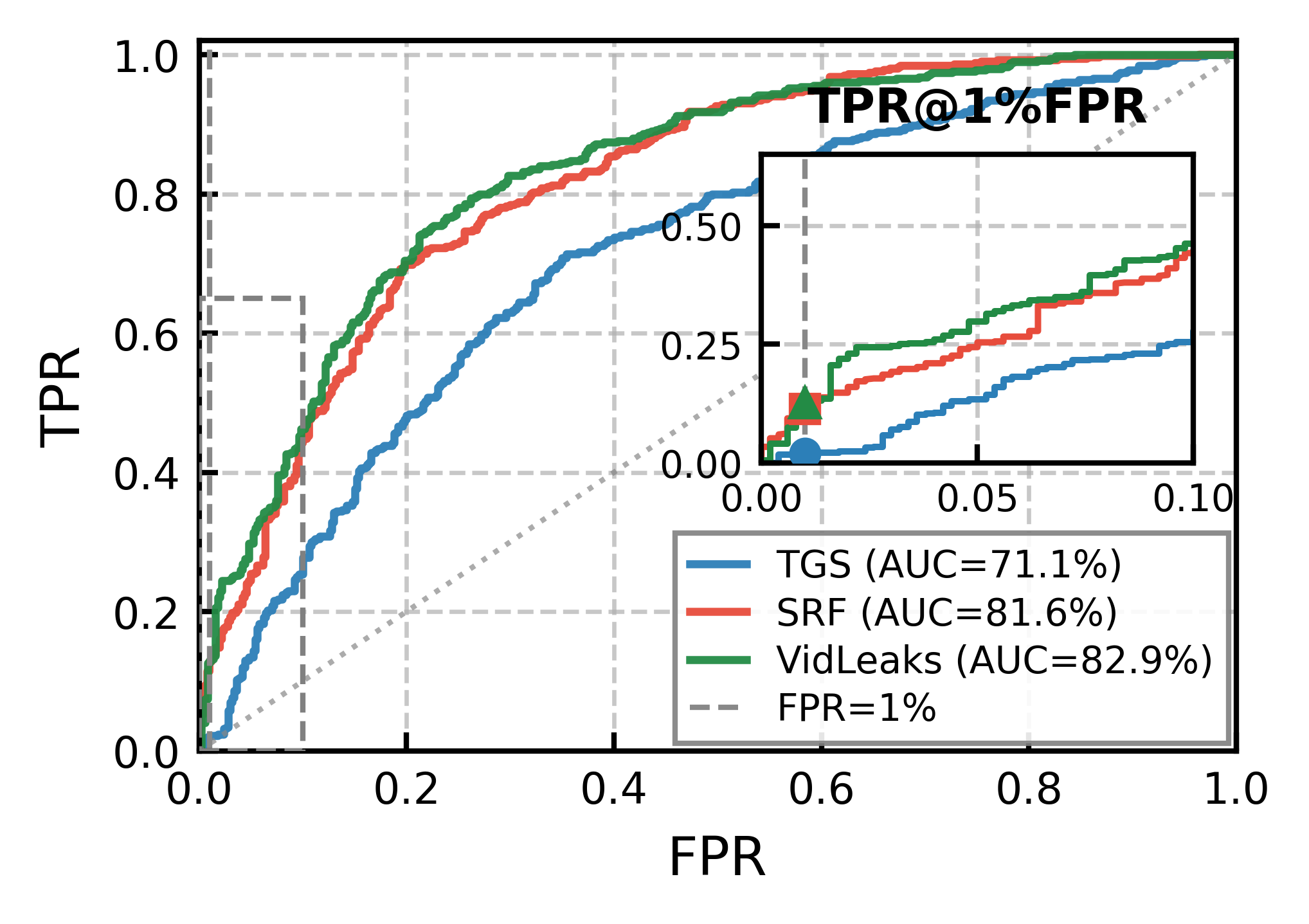}
    \subcaption{AnimateDiff}
    \label{fig:attack3_animatediff_roc}
  \end{subfigure}
  \hfill
  \begin{subfigure}[t]{0.32\textwidth}
    \centering
    \includegraphics[width=\linewidth]{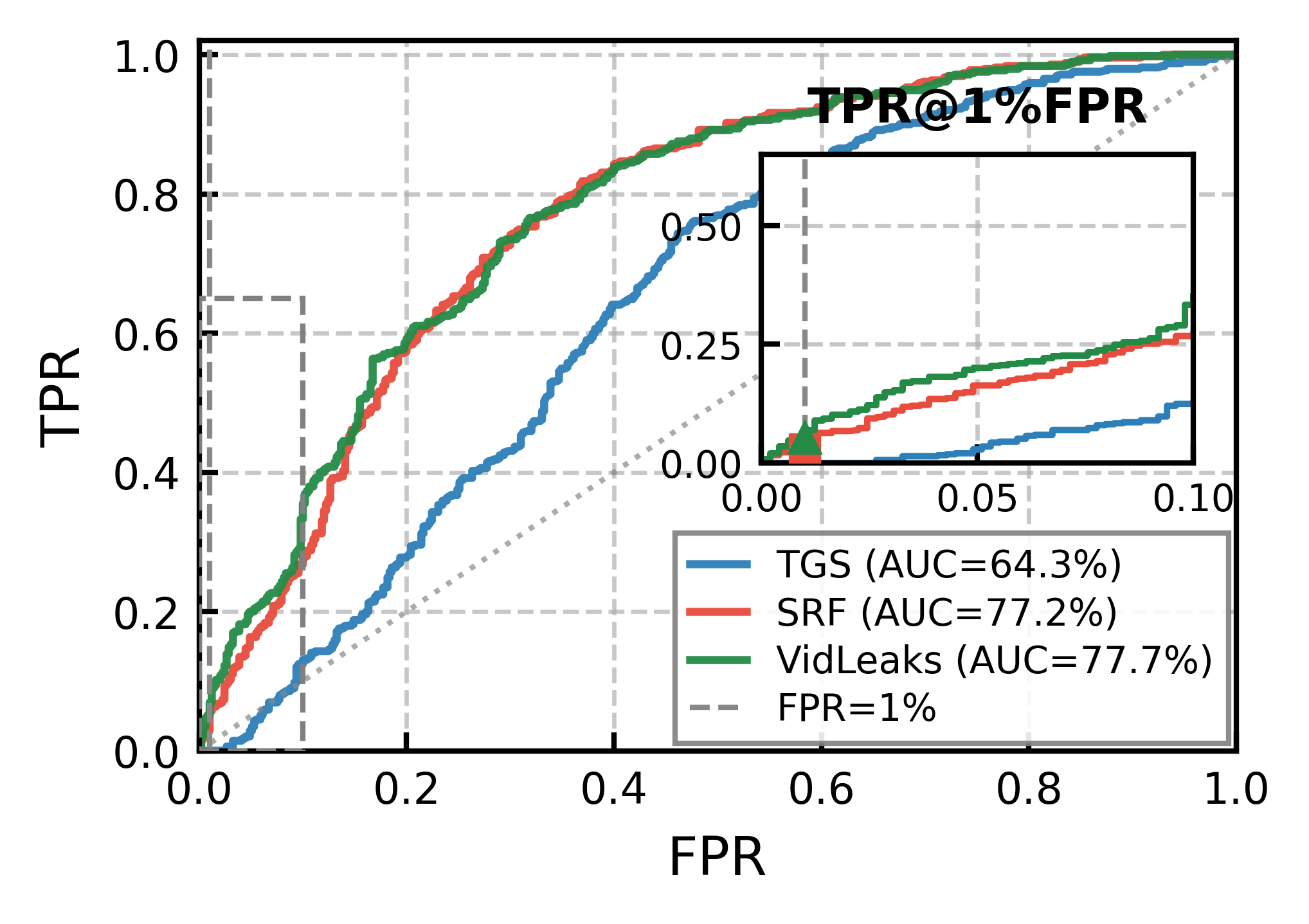}
    \subcaption{Mira}
    \label{fig:attack3_mira_roc}
  \end{subfigure}
  \hfill
  \begin{subfigure}[t]{0.32\textwidth}
    \centering
    \includegraphics[width=\linewidth]{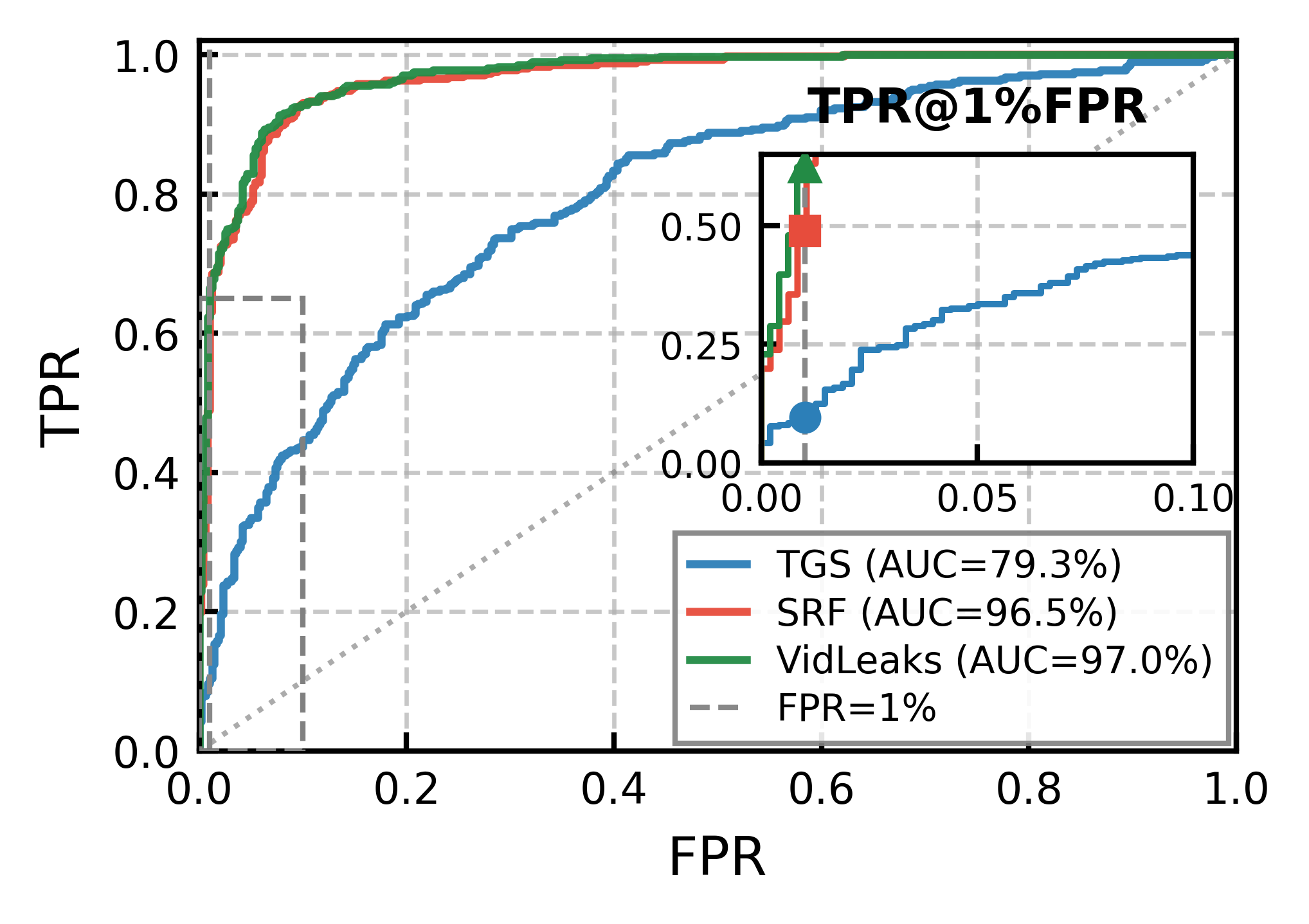}
    \subcaption{InstructVideo}
    \label{fig:attack3_instrcut_roc}  
  \end{subfigure}
  \caption{ROC curves for the query-only attack on different T2V models.}
  \label{fig:query_only_roc}
\end{figure*}

\paragraph{Temporal Dynamics of TGS Signal.}
To validate our TGS design, we analyze how its discriminative power evolves across the video timeline. For a generated video with $N$ frames, the $(N-1)$-dimensional instability vector (\autoref{eq:tgs_instab_vec}) is uniformly divided into three segments—Early, Middle, and Late—and segment-wise TGS scores are computed for evaluation.

\begin{figure}[t!]
  \centering
  \includegraphics[width=0.9\linewidth]{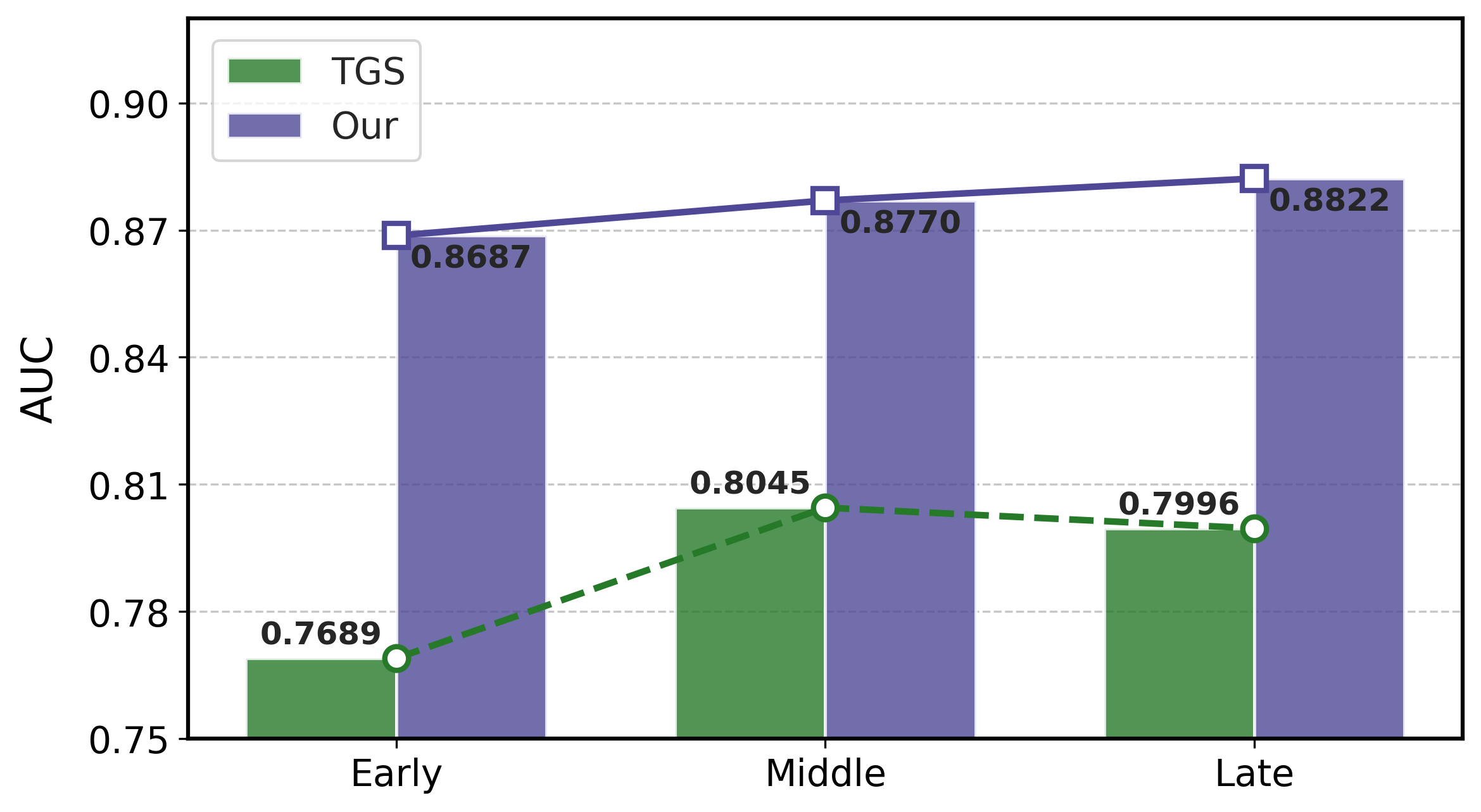}
  \caption{Analysis of TGS across temporal dimensions.} 
 \label{fig:ref_dim_impact}
\end{figure}

As shown in \autoref{fig:ref_dim_impact}, the membership signal tends to be weaker in early frames and becomes progressively stronger in the middle and late stages. This is consistent with our earlier visualization (\autoref{fig:core_signals_b}), where the instability gap between members and non-members widens as generation unfolds. These results empirically justify our per-dimension variance formulation. A naive “mean-then-std” alternative collapses temporal information, obscuring the stronger signals present in later frames. Direct comparison confirms this: our per-dimension design achieves an AUC of 82.63\%, significantly outperforming the 80.86\% of ``mean-then-std''. By preserving variance at each temporal step, TGS better captures the instability dynamics that expose membership leakage.

\section{Query-only Inference}
\label{sec:query_inference}
\subsection{Intuition and Assumptions}
\label{sec:query_only_intuition}
We consider the most restrictive yet practical threat model: \textbf{Query-Only Inference}. In this setting, the adversary operates under the zero-knowledge assumption, possessing neither a shadow dataset for training nor a reference set for calibration. Inference must be made solely from the signals derived for the target video, without any external calibration. This scenario simulates a real-world attacker with only black-box query access to the T2V model and the target video.

The key intuition is that our sparse-temporal signals are intrinsically discriminative. Although the global signal distributions are unknown, member samples are expected to exhibit both high reconstruction fidelity and high generative stability—a combination statistically rare among non-members. The attack therefore relies on unsupervised scoring and fusion of the two signals to infer membership.

\subsection{Attack Implementation}
\label{sec:query_only_methodology}
The query-only attack removes the need for calibration and relies solely on intrinsic scoring and unsupervised fusion of signals. The procedure consists of two stages: intrinsic signal scoring, and unsupervised fusion and inference (see Algorithm \ref{alg:query_only} in \autoref{app:algorithms}).

\paragraph{Intrinsic Signal Scoring.}
Without a reference distribution, we define intrinsic scores based on the inherent properties of the raw signals, following a unified ``higher is more member-like'' convention:

\begin{itemize}
    \item \textbf{SRF scoring:} $S_{SRF}$ already aligns with this convention, as higher values indicate membership. We thus use its raw value as the intrinsic score:
    \begin{equation}
    \label{eq:iscore_srf}
    \mathcal{S}_{SRF} = S_{SRF}
    \end{equation}
    \item \textbf{TGS scoring:} Since $S_{TGS}$ measures instability, lower values indicate membership. To align the directionality, we transform it into a stability score by inverting its value:
    \begin{equation}
    \label{eq:iscore_tgs}
    \mathcal{S}_{TGS} = 1 - S_{TGS}
    \end{equation}
\end{itemize}



\paragraph{Unsupervised Fusion and Inference.}
With the two intrinsic scores directionally aligned, we fuse them into a single membership score. Lacking data to learn optimal weights, we employ a pre-defined linear combination based on general heuristics about the relative importance of each signal:
\begin{equation}
\label{eq:fusion_query_only}
\mathcal{S}_{final} = w_{SRF} \cdot \mathcal{S}_{SRF} + w_{TGS} \cdot \mathcal{S}_{TGS}
\end{equation}
where $w_{SRF}$ and $w_{TGS}$ are pre-defined weights. This final score enables the adversary to rank candidate videos by membership likelihood. We evaluate $\mathcal{S}_{final}$ primarily with threshold-free metrics such as AUC, and report balanced accuracy without data-dependent calibration.

\begin{figure*}[t]
  \centering
  \includegraphics[width=0.98\linewidth]{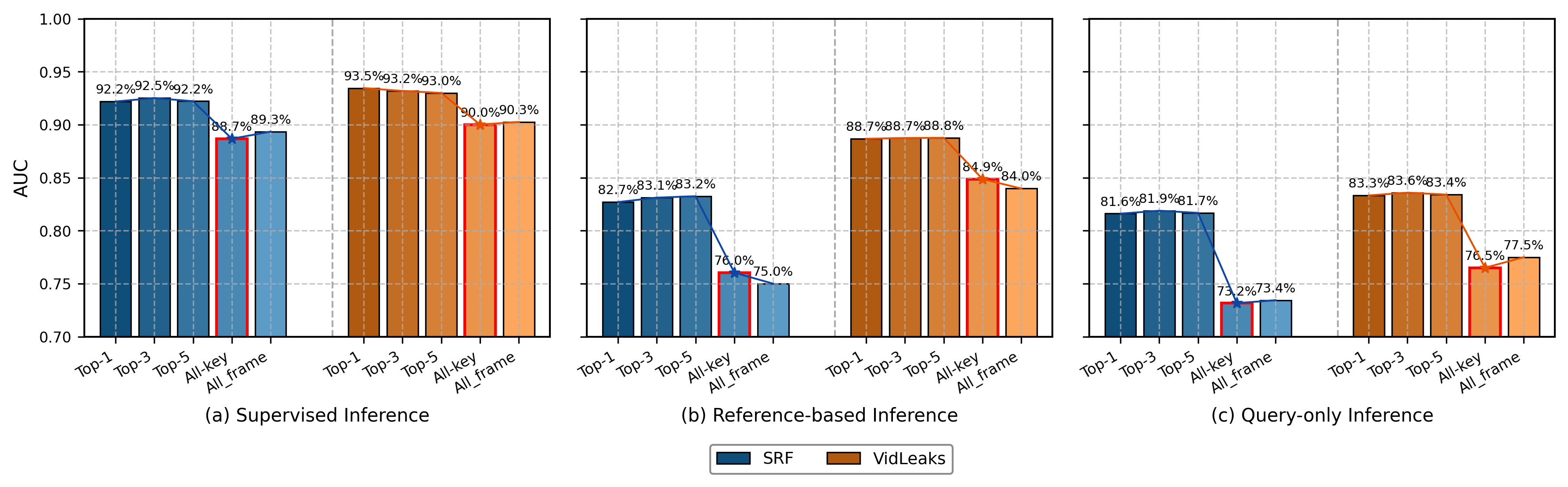}
  \caption{Impact of the Top-K strategy on SRF signal across three attack scenarios.}
  \label{fig:topk_ablation}
\end{figure*}

\subsection{Experimental Setting}
We again adopt the same models, datasets, and proxy captioning setup as in previous experiments, and report results with the same evaluation metrics (AUC, Balanced Accuracy, and TPR@1\%FPR).
In the query-only scenario, however, the adversary possesses neither a shadow dataset nor a non-member reference set. 
Inference must be made solely from the signals derived for the target video. Consequently, all member and non-member samples are directly used for evaluation.
Accordingly, we employ the intrinsic scoring and unsupervised fusion mechanism defined in \autoref{sec:query_only_methodology} using pre-defined weights ($w_{SRF} \approx w_{TGS}$). 
No external calibration or classifier training is required.

\subsection{Experimental Results}

\begin{table}[t!]
\centering
\caption{Performance of the query-only attack on different T2V models.}
\label{tab:query_only_results}
\resizebox{\linewidth}{!}{
\begin{tabular}{l l ccc}
\toprule
\textbf{Target Model} & \textbf{Method} 
& \textbf{AUC(↑)} & \textbf{TPR@1\%FPR(↑)} & \textbf{ACC(↑)} \\
\midrule
\multirow{3}{*}{AnimateDiff} 
  & SRF & 81.61\% & 11.40\% & 75.00\% \\
  & TGS & 71.07\% & 2.00\% & 68.00\% \\
  & \cellcolor{gray!10}\ourframework{} & \cellcolor{gray!10}\textbf{82.92\%} & \cellcolor{gray!10}\textbf{12.60\%} & \cellcolor{gray!10}\textbf{76.60\%} \\
\midrule
\multirow{3}{*}{Mira}        
  & SRF  & 77.18\% & 2.86\% & 72.45\% \\
  & TGS  & 64.27\% & 1.00\% & 64.18\% \\
  & \cellcolor{gray!10}\ourframework{} & \cellcolor{gray!10}\textbf{77.66\%} & \cellcolor{gray!10}\textbf{5.10\%} & \cellcolor{gray!10}72.55\% \\
\midrule
\multirow{3}{*}{InstructVideo} 
  & SRF  & 96.54\% & 48.88\% & 91.52\% \\
  & TGS  & 79.32\% & 9.68\% & 72.55\% \\
  & \cellcolor{gray!10}\ourframework{} & \cellcolor{gray!10}\textbf{97.01\%} & \cellcolor{gray!10}\textbf{62.28\%} & \cellcolor{gray!10}\textbf{91.80\%} \\
\bottomrule
\end{tabular}}
\end{table}

\paragraph{Overall Performance.}
\autoref{tab:query_only_results} shows that our attack remains highly effective even in the most restrictive \textit{query-only} setting. Without access to any reference distribution or labeled data, our method still achieves an AUC of 82.92\% on AnimateDiff and 77.66\% on Mira, while reaching a striking 97.01\% on InstructVideo. These results demonstrate that the sparse-temporal artifacts of memorization are sufficiently strong to be detected directly from raw signals, with no external calibration. The ROC curves in \autoref{fig:query_only_roc} further confirm that our fused model consistently dominates the single-signal baselines, particularly in the low-FPR region where practical risks are most relevant.

\paragraph{Intrinsic Signal Power.}
The query-only results also provide insight into the relative strength of our signals. As shown in \autoref{tab:query_only_results}, SRF consistently outperforms TGS when used alone, indicating that sparse reconstruction fidelity serves as a more absolute, calibration-free indicator of memorization. By contrast, temporal stability (TGS) is more sensitive to the lack of reference normalization, which explains its weaker standalone performance in this setting. Nevertheless, fusing SRF and TGS yields the strongest overall performance across all models, confirming that their complementary nature persists even under zero-knowledge assumptions. This highlights the robustness of our \ourframework{} and underscores the severe privacy risks posed by T2V models.

\section{Ablation and Robustness Analysis}
\label{ablation}

\subsection{Effectiveness of the Top-K Strategy}
\label{ablation: top-k}
Our Sparse Reconstruction Fidelity (SRF) is designed as a matched filter over \emph{key anchors}, averaging similarities only over the Top-K most similar keyframes to combat content sparsity. We evaluate how $K$ affects performance by comparing $K\!\in\!\{1,3,5\}$ against two averaging baselines: \textit{All-key} (average over all keyframes, i.e., no Top-K filtering) and \textit{All-frame} (average over all frames). \autoref{fig:topk_ablation} reports AUC for both SRF-only and our fused attack (\ourframework{}) across the three threat models.

The results reveal two consistent trends: (1) Performance is consistently stable for $K\!\in\!\{1,3,5\}$ in all scenarios and for both SRF and \ourframework{}, indicating that SRF is insensitive to the precise choice of a small $K$ so long as it focuses on the most relevant anchors; and (2) Removing the Top-K filter causes a clear drop: both \textit{All-key} and \textit{All-frame} substantially underperform Top-K variants across settings. This validates our core hypothesis from \autoref{sec:spatial_probe}: averaging over many non-memorized (generalized) frames dilutes the sparse membership signal carried by a few memorized key anchors. 

In summary, the Top-$K$ mechanism is essential for isolating spatial memorization; a small $K$ (e.g., $K\!=\!3$) provides a robust choice that consistently benefits SRF and, by extension, \ourframework{} across all threat models.

\begin{figure*}[!t]
  \centering
  \includegraphics[width=0.97\linewidth]{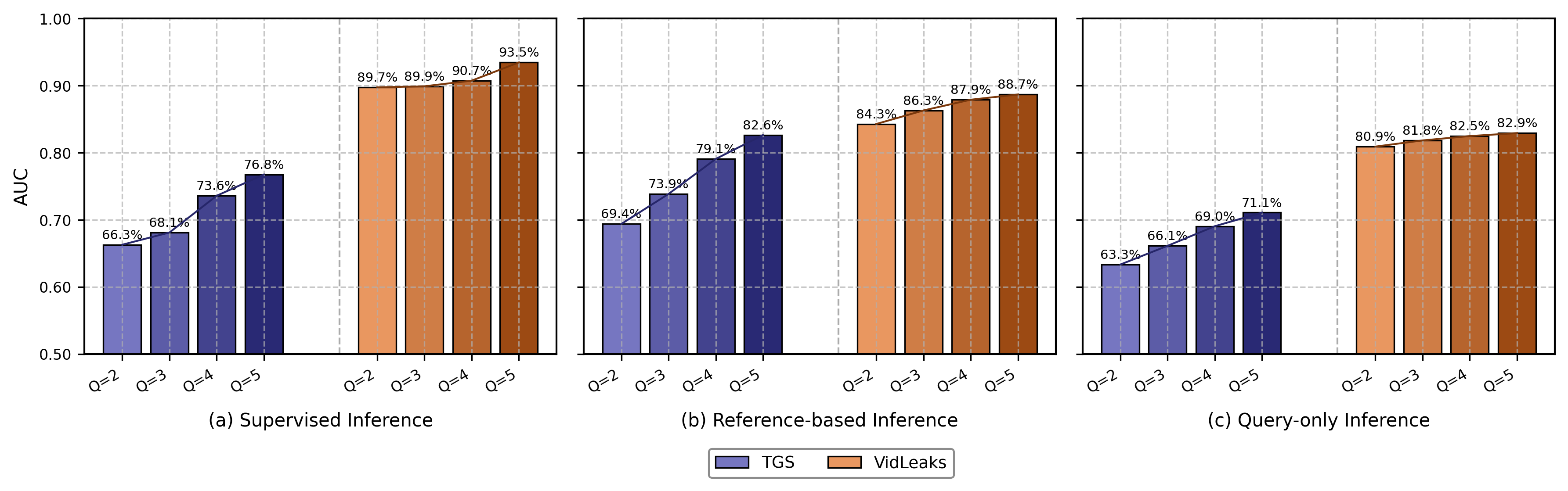}
\caption{Impact of query count ($Q$) on TGS signal across three attack scenarios.}
\label{fig:ablation_multiQ}
\end{figure*}

\begin{table*}[t]
\renewcommand{\arraystretch}{1.5}
\centering
\caption{Comparison of TGS signal against alternative temporal signals across three attack scenarios.}
\label{tab:alternative_motion}
\resizebox{0.96\linewidth}{!}{
\begin{tabular}{lccccccccc}
\toprule
\multirow{2}{*}{\centering \textbf{Signal Type}} & \multicolumn{3}{c}{\textbf{Supervised}} & \multicolumn{3}{c}{\textbf{Reference-based}} & \multicolumn{3}{c}{\textbf{Query-only}} \\
\cmidrule(lr){2-4} \cmidrule(lr){5-7} \cmidrule(lr){8-10}
 & \textbf{AUC(↑)} & \textbf{TPR@1\%FPR(↑)} & \textbf{ACC(↑)} & \textbf{AUC(↑)} & \textbf{TPR@1\%FPR(↑)} & \textbf{ACC(↑)} & \textbf{AUC(↑)} & \textbf{TPR@1\%FPR(↑)} & \textbf{ACC(↑)} \\
\midrule
\textbf{Subject Consistency} & 65.22\% & 0.84\% & 62.12\% & 74.19\% & 4.20\% & 68.13\% & 63.17\% & 1.20\% & 59.40\% \\
\textbf{Temporal Jitter} & 68.59\% & 1.84\% & 65.38\% & 67.42\% & 5.04\% & 67.26\% & 60.33\% & 1.00\% & 59.80\% \\
\cellcolor{gray!10}\textbf{TGS} &\cellcolor{gray!10}\textbf{76.75\%} & \cellcolor{gray!10}\textbf{2.52\%} & \cellcolor{gray!10}\textbf{72.03\%} & \cellcolor{gray!10}\textbf{82.63\%} & \cellcolor{gray!10}\textbf{5.88\%} & \cellcolor{gray!10}\textbf{77.65\%} & \cellcolor{gray!10}\textbf{71.07\%} & \cellcolor{gray!10}\textbf{2.00\%} & \cellcolor{gray!10}\textbf{68.00\%} \\
    
\bottomrule
\end{tabular}
}
\end{table*}

\subsection{Effectiveness of the Multi-Query Strategy}
\label{ablation:multi-q}
Our Temporal Generative Stability (TGS) signal relies on variance across multiple generations to capture determinism in temporal dynamics. To validate this design, we analyze the effect of the number of queries ($Q$) on attack performance. \autoref{fig:ablation_multiQ} reports AUC scores for both the TGS-only attack and our fused method as $Q$ increases from 2 to 5.

The results show a consistent and substantial performance gain with more queries. For instance, in the supervised setting, TGS improves from 66.3\% ($Q$=2) to 76.8\% ($Q$=5), while our fused method rises from 89.7\% to 93.5\%. Similar monotonic improvements are observed in the reference-based and query-only scenarios, with performance stabilizing when $Q \geq 4$. This trend confirms our hypothesis: a single generation cannot reliably reflect temporal stability, while aggregating across multiple runs yields a robust estimate of generative variance.

Overall, the Multi-Query design is essential for extracting strong temporal memorization signals, making TGS both effective and stable across different inference settings.

\subsection{Comparison with Alternative Temporal Signals}
\label{ablation: motion}

To validate our design for Temporal Generative Stability (TGS), we compare it with two intuitive alternatives for capturing temporal dynamics: 
(1) \textbf{Subject Consistency}, a semantic-level metric that measures whether a subject’s appearance remains consistent across frames using DINO features~\cite{caron2021emerging}; and 
(2) \textbf{Temporal Jitter}, a low-level motion metric that quantifies inter-frame pixel changes via RAFT optical flow~\cite{teed2020raft}. 
For fairness, both alternatives are adapted into instability signals using the same multi-query formulation as TGS.   

The results in \autoref{tab:alternative_motion} clearly show that:  
(1) \textit{Low-level motion (Temporal Jitter) fails as a membership signal}, with AUC dropping to only 60.33\% in the query-only scenario, indicating that pixel-level dynamics are too stochastic to separate members from non-members.  
(2) \textit{Subject Consistency has moderate discriminative power}, but remains substantially weaker than TGS across all settings. In the critical reference-based scenario, for example, it achieves only 74.19\% AUC compared to 82.63\% for TGS.  
(3) \textit{TGS consistently outperforms both alternatives}, achieving the best results across all metrics and scenarios.  

These findings empirically validate our core intuition: focusing solely on a subject introduces noise from its legitimate motion, while low-level pixel dynamics are inherently unstable. By contrast, TGS leverages the stability of the holistic scene, anchored by relatively static elements, to provide a robust and highly discriminative temporal signal for membership inference.  


\begin{table*}[t]
\renewcommand{\arraystretch}{1.5}
\centering
\caption{Impact of caption source and quality on attack performance across three attack scenarios.}
\label{tab:captioning_impact}
\resizebox{\linewidth}{!}{
\begin{tabular}{lccccccccc}
\toprule
\multirow{2}{*}{\centering \textbf{Prompt Type}} & \multicolumn{3}{c}{\textbf{Supervised}} & \multicolumn{3}{c}{\textbf{Reference-based}} & \multicolumn{3}{c}{\textbf{Query-only}} \\
\cmidrule(lr){2-4} \cmidrule(lr){5-7} \cmidrule(lr){8-10}
 & \textbf{AUC(↑)} & \textbf{TPR@1\%FPR(↑)} & \textbf{ACC(↑)} & \textbf{AUC(↑)} & \textbf{TPR@1\%FPR(↑)} & \textbf{ACC(↑)} & \textbf{AUC(↑)} & \textbf{TPR@1\%FPR(↑)} & \textbf{ACC(↑)} \\
\midrule
\textbf{Ground-Truth} & 89.06\% & 32.77\% & 83.24\% & 86.99\% & 10.84\% & 79.81\% & 80.67\% & 8.60\% & 74.70\% \\

\textbf{Doubao} & 98.06\% & 62.57\% & 93.39\% & 90.84\% & 34.45\% & 84.59\% & 85.25\% & 31.00\% & 77.50\% \\
\textbf{30\% Word Dropout } & 90.34\% & 34.35\% & 84.12\% & 88.55\% & 32.77\% & 81.62\% & 82.53\% & 12.80\% & 74.70\% \\
\cellcolor{gray!10}\textbf{Gemini Pro (default)} &\cellcolor{gray!10}93.46\% & \cellcolor{gray!10}49.58\% & \cellcolor{gray!10}87.58\% & \cellcolor{gray!10}88.68\% & \cellcolor{gray!10}27.73\% & \cellcolor{gray!10}82.84\% & \cellcolor{gray!10}82.92\% & \cellcolor{gray!10}12.60\% & \cellcolor{gray!10} 76.60\% \\
\bottomrule
\end{tabular}
}
\end{table*}

\begin{table}[t!]
\centering
\caption{Comparison with static similarity baselines across three attack scenarios.}
\label{tab:similarity_metrics_results}
\resizebox{\linewidth}{!}{
\begin{tabular}{lccc}
\toprule
\textbf{Method} & \textbf{Supervised} & \textbf{Reference-based} & \textbf{Query-only} \\
\midrule
Frame-wise & 89.3\% & 75.0\% & 73.4\% \\
Video-level & 57.2\% & 53.0\% & 47.5\% \\
\rowcolor{gray!10}
\textbf{\ourframework{}} & \textbf{93.5\%} & \textbf{88.7\%} & \textbf{82.9\%} \\
\bottomrule
\end{tabular}}
\end{table}

\subsection{Comparison with Static Similarity Baselines}
\label{sec:similarity metrics}

To assess the applicability of static similarity metrics in the video-based MIA setting, we compare \ourframework{} against two intuitive baselines that are derived from image-based MIAs: (1) \textit{Frame-wise CLIP Similarity}: Following~\cite {wu2022membership}, we compute CLIP similarity for each generated frame against all frames of the target video and average the results. This represents the most straightforward extension of image-based MIA to videos. (2) \textit{Video-level Similarity}: We compute cosine similarity between VideoCLIP~\cite{wang2024internvideo2} embeddings of the target and generated videos, capturing holistic video semantics.

As shown in \autoref{tab:similarity_metrics_results}, both baselines substantially underperform our \ourframework{} across all threat scenarios. Frame-wise similarity achieves only 73.4\% AUC in the query-only setting (vs. 82.9\% for \ourframework{}), while video-level similarity yields near-random performance (AUC ~47.5-57.2\%).  These results show that static similarity metrics fail to handle the sparsity and temporal dynamics of video memorization. Our SRF and TGS, however, specifically target and capture these sparse-temporal leakage signals from T2V models.


\subsection{Impact of Caption Source and Quality}
\label{sec:ablation_captioning}
Our attack pipeline assumes a practical setting where the adversary only has access to the target video and thus must rely on a public captioning tool to generate a \textbf{proxy caption}. A natural question is how such proxy captions compare with the \textbf{ground-truth captions} used in training, and whether the attack is robust to variations in caption quality. 
To examine this, we evaluate our attack across all three inference scenarios using multiple caption sources: (1) ground-truth captions from the original datasets, (2) proxy captions generated by Gemini Pro~\cite{GoogleAIStudio}, (3) proxy captions generated by an alternative model (Doubao~\cite{VolcEngineDoubao}), and (4) degraded captions obtained by applying 30\% random word dropout to Gemini Pro's outputs.

As shown in \autoref{tab:captioning_impact}, proxy captions not only avoid degrading performance but in fact \emph{often outperform} ground-truth captions. For example, in the supervised setting, AUC increases from 89.06\% with ground-truth captions to 93.46\% with Gemini Pro, and further to 98.06\% with Doubao. Similar improvements appear under the reference-based and query-only scenarios. This counter-intuitive result may arise from two factors: (1) proxy captions produced by modern captioning models are typically more descriptive and semantically detailed, which forces the T2V model to reconstruct fine-grained spatial and temporal anchors, thereby amplifying SRF and TGS differences between members and non-members; and (2) in contrast, ground-truth captions from datasets such as WebVid-10M are often short or weakly descriptive. Although they were used during training, their limited expressiveness may fail to fully activate the model’s memorized patterns. 

Importantly, even under substantial caption degradation (30\% random word dropout), the attack maintains strong performance (e.g., 90.34\% AUC in the supervised setting), demonstrating robustness to imperfect caption quality. These findings reinforce the practicality of our video-only pipeline and show that advanced external captioners can, perhaps unexpectedly, yield stronger membership inference attacks than the ground-truth captions themselves.

\begin{table*}[t]
\renewcommand{\arraystretch}{1.5}
\centering
\caption{Performance under API-Level perturbation defenses across three attack scenarios.}
\label{tab:defense_robustness}
\resizebox{\linewidth}{!}{
\begin{tabular}{lccccccccc}
\toprule
\multirow{2}{*}{\centering \textbf{Defense Setting}} 
& \multicolumn{3}{c}{\textbf{Supervised}} 
& \multicolumn{3}{c}{\textbf{Reference-based}} 
& \multicolumn{3}{c}{\textbf{Query-only}} \\
\cmidrule(lr){2-4} 
\cmidrule(lr){5-7} 
\cmidrule(lr){8-10}
& \textbf{AUC(↑)} & \textbf{TPR@1\%FPR(↑)} & \textbf{ACC(↑)}
& \textbf{AUC(↑)} & \textbf{TPR@1\%FPR(↑)} & \textbf{ACC(↑)}
& \textbf{AUC(↑)} & \textbf{TPR@1\%FPR(↑)} & \textbf{ACC(↑)} \\
\midrule

\textbf{cfg\_scale $\pm 0.5$} 
& 93.48\% & 52.94\% & 87.48\%
& 89.97\% & 19.33\% & 81.84\%
& 83.96\% & 9.80\%  & 76.90\% \\

\textbf{inference\_steps $\pm 1$} 
& 92.25\% & 57.14\% & 84.51\%
& 89.82\% & 21.01\% & 82.50\%
& 82.53\% & 8.40\%  & 75.70\% \\

\cellcolor{gray!10}\textbf{\ourframework{}}
& \cellcolor{gray!10}93.46\% & \cellcolor{gray!10}49.58\% & \cellcolor{gray!10}87.58\%
& \cellcolor{gray!10}88.68\% & \cellcolor{gray!10}27.73\% & \cellcolor{gray!10}82.84\%
& \cellcolor{gray!10}82.92\% & \cellcolor{gray!10}12.60\% & \cellcolor{gray!10}76.60\% \\

\bottomrule
\end{tabular}
}
\end{table*}

\subsection{Dataset Distribution Shift Analysis}
\label{sec:bias_analysis}

In our evaluation, member samples originate from WebVid-10M (AnimateDiff, InstructVideo) or MiraData (Mira), while all non-member samples are drawn from Panda-70M. Although these datasets share broad semantic domains, they differ in collection pipelines, visual quality, and content composition, which may induce dataset-level distributional shifts. It is therefore important to assess whether the signals exploited by our attack reflect genuine model memorization or merely these distributional differences.

To isolate the effect of dataset-level distributional shift, we employ a \emph{blind classifier}: an MLP trained solely on VideoCLIP embeddings extracted from raw videos (100 epochs, learning rate $1\times10^{-5}$), without using any generated outputs or temporal signals from the target T2V models. This reflects an upper bound on the separability attributable purely to dataset statistics. As shown in~\autoref{tab:blind_classifier}, the blind classifier reveals varying degrees of distribution shift across models. AnimateDiff shows the weakest shift (AUC 68.39\%), InstructVideo exhibits a moderate shift (AUC 75.27\%), likely due to its animal-focused fine-tuning, and Mira demonstrates the strongest shift (AUC 82.05\%), indicating a more pronounced mismatch between MiraData and Panda-70M. Notably, despite these shifts, the blind classifier’s discriminative power remains limited, particularly in the low-FPR regime.

Crucially, this distributional shift pattern does not align with the performance of \ourframework{}. Our attack achieves substantially higher AUC on InstructVideo and AnimateDiff (up to 98.04\%) despite their smaller dataset shifts, while its performance on Mira (AUC 87.45\%) remains lower even though the distributional shift is greatest. Moreover, the blind classifier yields only weak high-confidence inference (TPR@1\%FPR 2.38\%--7.00\%), far below the supervised performance of \ourframework{} (TPR@1\%FPR 12.00\%--49.58\%). These results, combined with the distinct SRF and TGS patterns of member videos (\autoref{fig:core_signals}), demonstrate that dataset-level distributional bias alone cannot account for our results. Instead, \ourframework{} exposes genuine sparse-temporal memorization effects inherent to the generative behavior of T2V models. 

\begin{table}[t!]
\centering
\caption{Comparison between a blind dataset-based classifier and \ourframework{} under the supervised setting. }
\label{tab:blind_classifier}
\resizebox{\linewidth}{!}{
\begin{tabular}{lcccc}
\toprule
\textbf{Target Model} & \textbf{Method} & \textbf{AUC} & \textbf{TPR@1\%FPR} & \textbf{ACC} \\
\midrule
\multirow{2}{*}{AnimateDiff}
  & Blind Classifier & 68.39\% & 3.36\% & 64.10\% \\
  &\cellcolor{gray!10}\textbf{\ourframework{}}  &\cellcolor{gray!10}\textbf{93.46\%}  &\cellcolor{gray!10}\textbf{49.58\%}  &\cellcolor{gray!10}\textbf{87.58\%} \\
\addlinespace
\multirow{2}{*}{Mira}
  & Blind Classifier & 82.05\% & 7.00\% & 76.00\% \\
  &\cellcolor{gray!10}\textbf{\ourframework{}}  &\cellcolor{gray!10}\textbf{87.45\%}  &\cellcolor{gray!10}\textbf{12.00\%}  &\cellcolor{gray!10}\textbf{82.00\%} \\
\addlinespace
\multirow{2}{*}{InstructVideo}
  & Blind Classifier & 75.27\% & 2.38\% & 64.29\% \\
   &\cellcolor{gray!10}\textbf{\ourframework{}}  &\cellcolor{gray!10}\textbf{98.04\%}  &\cellcolor{gray!10}\textbf{45.00\%}  &\cellcolor{gray!10}\textbf{94.77\%} \\
\bottomrule
\end{tabular}}
\end{table}




\section{Discussion}
\label{sec:discussion}

Our study demonstrates that state-of-the-art T2V models are highly vulnerable to membership inference attacks through their sparse-temporal memorization. In this section, we discuss the broader implications of our findings, potential countermeasures, and the limitations of our study.

\paragraph{Summary of Key Findings.}
Our work provides three major insights into how T2V models memorize. First, we show that memorization is inherently dual-faceted: models tend to preserve \emph{salient visual anchors} (captured by SRF) and \emph{stable temporal dynamics} (captured by TGS). Second, we establish that these two signals are complementary rather than redundant—their fusion consistently amplifies attack power across architectures, datasets, and threat models. Third, and most critically, we reveal that these memorization artifacts are so strong that they can be reliably exploited even in the strictest \textbf{query-only} setting—without access to member data, non-member baselines, or model internals. 
This suggests that the membership leakage is not merely a statistical anomaly but a fundamental vulnerability of current T2V models, posing a severe and practical threat. 

\paragraph{Countermeasures and Defenses.}
Mitigating our sparse-temporal MIA is a non-trivial challenge, as the exploited signals are inherently tied to the generative objectives of T2V models. Accordingly, effective mitigation is likely to require interventions that reduce memorization during training and data construction. Below, we outline two potential defense directions.
(1) \textbf{Training-Time Defenses}. The fundamental protections are integrated during model optimization. For example, Differential Privacy (DP)~\cite{abadi2016deep} offers strong theoretical guarantees but remains impractical for large-scale generative models due to prohibitive costs. Alternatively, targeted regularization strategies could be designed to penalize over-deterministic temporal generation or excessively faithful reconstruction of training samples. (2) \textbf{Data Pre-processing}. Since memorization often stems from unique training examples, thorough filtering and de-duplication of pre-training corpora~\cite{lee2021deduplicating} may help reduce leakage risk. However, this comes at the potential cost of diminished data diversity and downstream performance.

Given that training-time modifications and large-scale data curation may be costly to deploy, we further evaluate a lightweight \textbf{API-level defense} that can be readily applied to existing systems.
Specifically, we consider simple parameter randomization, where each query introduces small random jitter to key sampling parameters (\textit{cfg\_scale}~$\pm 0.5$ and \textit{inference\_steps}~$\pm 1$). 
As shown in~\autoref{tab:defense_robustness}, \ourframework{} remains highly robust across all threat models: in supervised settings, AUC varies by less than $1.5\%$, and in the query-only setting performance remains above $82.5\%$.
These results indicate that such lightweight perturbations do not meaningfully suppress the sparse-temporal signals that drive the attack, underscoring the limitations of simple API-level defenses.

Overall, our analysis suggests that mitigating membership leakage in modern T2V systems will likely require a combination of deeper training-time interventions, careful data curation, and system-level safeguards.
Designing defenses that effectively suppress sparse-temporal memorization while preserving generation quality and usability remains an important and open challenge for future research.

\paragraph{Limitations and Future Work.}
Despite its strengths, our study has several limitations that suggest directions for future research. First, our evaluation focuses on three representative T2V models, which cover the major architectural paradigms but do not exhaust the rapidly evolving landscape of video generation. 
Extending the analysis to newer or larger-scale systems will be critical for validating the generality of our findings. 
Second, although we demonstrate that proxy captions generated by a public captioner are not only practical but often more effective than ground-truth captions (\autoref{sec:ablation_captioning}), our exploration of this dependency is not exhaustive. A deeper investigation into how different captioning strategies or language models influence membership inference remains an important open question. 
Finally, while we have outlined potential defenses, their concrete design, implementation, and rigorous evaluation are non-trivial. Developing privacy-preserving training and inference mechanisms that specifically mitigate the sparse-temporal leakage channels we identify represents an important and challenging direction for future work. 

\section{Conclusion}
\label{sec:conclusion}
In this paper, we present the first systematic study of membership inference attacks against modern text-to-video models. We identify two fundamental challenges unique to this domain—the sparsity of content memorization and the dynamics of temporal memorization—and propose the sparse-temporal MIA framework (\ourframework{}) to address them. Our framework introduces two complementary signals: Sparse Reconstruction Fidelity (SRF) for detecting salient visual anchors and Temporal Generative Stability (TGS) for capturing stable temporal evolution.

Through extensive experiments across three diverse T2V models and under multiple threat models, we have demonstrated that these systems are highly vulnerable to membership inference. Notably, our attack remains effective even in the strict zero-knowledge, query-only setting, revealing a severe and practical privacy risk.  
Our study provides the first concrete evidence that T2V models leak significant membership information through both sparse fidelity and temporal dynamics, underscoring the urgent need for targeted defenses to safeguard data privacy and content ownership in the era of generative video. 

\appendix
\section*{Acknowledgments} 
This work was supported by National Natural Science Foundation of China under Grant (No. 62372268, No. 62502276), Key R\&D Program of Shandong Province, China (No. 2024CXGC010114, No. 2025CXPT085), and the Postdoctoral Innovation Program of Shandong Province, China (No. SDCX-ZG-202503030).

\section*{Ethical Considerations}
This work presents the first systematic analysis of membership inference attacks against text-to-video (T2V) models. As security research that exposes system vulnerabilities, we fully acknowledge its dual-use potential and have carefully considered its ethical implications.

\paragraph{Stakeholders and Potential Impact.}
Our work primarily involves three stakeholder groups: 
(1) \textit{T2V Model Developers and Companies}. Our research reveals previously unexplored privacy risks in their systems, enabling a clearer understanding of when and why MIAs succeed. This knowledge can support the development of defenses that reduce unintended memorization and improve privacy guarantees. 
(2) \textit{Content Creators and Data Subjects}. The methodology clarifies the technical feasibility and limitations of using MIA to assess whether specific content may have been used during training. While this does not establish a definitive or legally robust auditing mechanism, it helps delineate the evidentiary strength and constraints of such approaches. Notably, improvements in model defenses informed by this line of research may simultaneously reduce the effectiveness of membership inference for external auditing, reflecting an inherent tension between privacy protection and auditability. 
(3) \textit{The Research and Policy Community}. We provide crucial, empirical insights into video-level privacy risks, informing the development of ethical guidelines and technical standards for generative video technologies. A concurrent risk is that the publication could inform malicious actors.

\paragraph{Mitigations and Responsible Disclosure.} 
We use only publicly released datasets under their respective licenses and do not attempt to identify individuals or recover video content. The attack is designed solely to reveal training data membership and is evaluated under progressively restrictive threat models. We provide methodological detail for scientific validation but intentionally avoid releasing turnkey attack tools, consistent with responsible disclosure principles.

\paragraph{Justification and Risk-Benefit Balance.}
Our work follows the preventive security research paradigm: proactively identifying privacy vulnerabilities in T2V models allows defenses to be developed before potential exploitation. Although the methodology could in principle be misused for membership inference, the risk is limited and further mitigated by our responsible disclosure practices. By contrast, concealing such vulnerabilities would leave data subjects exposed without impeding malicious discovery. Transparent disclosure provides clear benefits—strengthening accountability, informing risk assessment, and motivating privacy-preserving design. Therefore, responsible publication with ethical reflection serves the net interests of the security community and society. 

\section*{Open Science}

We follow open science principles to promote transparency and reproducibility. 
We provide access to the codebase implementing our attack pipeline, along with documentation describing how to set up the environment, obtain the publicly available datasets used in our experiments, and reproduce the main results reported in the paper. 
Because the raw video data are large and subject to license restrictions, we do not redistribute them directly; instead, we include metadata and scripts for retrieving the videos from their original public sources. All materials used in this paper — including code, configuration files, and representative result files — are available in a public repository: \url{https://zenodo.org/records/17972831.}

\cleardoublepage
\bibliographystyle{plain}
\bibliography{references}

@inproceedings{ho2020denoising,
  title={Denoising diffusion probabilistic models},
  author={Ho, Jonathan and Jain, Ajay and Abbeel, Pieter},
  booktitle={Advances in Neural Information Processing Systems},
  volume={33},
  pages={6840--6851},
  year={2020}
}

@inproceedings{rombach2022high,
  title={High-resolution image synthesis with latent diffusion models},
  author={Rombach, Robin and Blattmann, Andreas and Lorenz, Dominik and Esser, Patrick and Ommer, Bj{\"o}rn},
  booktitle={Proceedings of the IEEE/CVF conference on computer vision and pattern recognition},
  pages={10684--10695},
  year={2022}
}

@article{song2020score,
  title={Score-based generative modeling through stochastic differential equations},
  author={Song, Yang and Sohl-Dickstein, Jascha and Kingma, Diederik P and Kumar, Abhishek and Ermon, Stefano and Poole, Ben},
  journal={arXiv preprint arXiv:2011.13456},
  year={2020}
}

@misc{sora2024,
  author       = {OpenAI},
  title        = {Sora},
  year         = {2024},
  howpublished = {\url{https://openai.com/index/sora/}},
}

@misc{kling2024,
  author       = {{Kwai}},
  title        = {Kling},
  howpublished = {\url{https://kling.kuaishou.com}},
  year         = {2024},
  note         = {Accessed: Aug. 2025}
}

@misc{luma2024dream,
  author       = {{Luma}},
  title        = {Luma Dream Machine},
  howpublished = {\url{https://lumalabs.ai/dream-machine}},
  year         = {2024},
}

@misc{runway2024gen3,
  author       = {{Runway}},
  title        = {Gen-3},
  howpublished = {\url{https://runwayml.com/blog/introducing-gen-3-alpha}},
  year         = {2024},
}

@inproceedings{sun2025t2v,
  title={T2v-compbench: A comprehensive benchmark for compositional text-to-video generation},
  author={Sun, Kaiyue and Huang, Kaiyi and Liu, Xian and Wu, Yue and Xu, Zihan and Li, Zhenguo and Liu, Xihui},
  booktitle={Proceedings of the Computer Vision and Pattern Recognition Conference},
  pages={8406--8416},
  year={2025}
}

@article{wang2025lavie,
  title={Lavie: High-quality video generation with cascaded latent diffusion models},
  author={Wang, Yaohui and Chen, Xinyuan and Ma, Xin and Zhou, Shangchen and Huang, Ziqi and Wang, Yi and Yang, Ceyuan and He, Yinan and Yu, Jiashuo and Yang, Peiqing and others},
  journal={International Journal of Computer Vision},
  volume={133},
  number={5},
  pages={3059--3078},
  year={2025},
  publisher={Springer}
}

@article{blattmann2023stable,
  title={Stable video diffusion: Scaling latent video diffusion models to large datasets},
  author={Blattmann, Andreas and Dockhorn, Tim and Kulal, Sumith and Mendelevitch, Daniel and Kilian, Maciej and Lorenz, Dominik and Levi, Yam and English, Zion and Voleti, Vikram and Letts, Adam and others},
  journal={arXiv preprint arXiv:2311.15127},
  year={2023}
}

@article{yang2024cogvideox,
  title={Cogvideox: Text-to-video diffusion models with an expert transformer},
  author={Yang, Zhuoyi and Teng, Jiayan and Zheng, Wendi and Ding, Ming and Huang, Shiyu and Xu, Jiazheng and Yang, Yuanming and Hong, Wenyi and Zhang, Xiaohan and Feng, Guanyu and others},
  journal={arXiv preprint arXiv:2408.06072},
  year={2024}
}

@article{fan2025vchitect,
  title={Vchitect-2.0: Parallel Transformer for Scaling Up Video Diffusion Models},
  author={Fan, Weichen and Si, Chenyang and Song, Junhao and Yang, Zhenyu and He, Yinan and Zhuo, Long and Huang, Ziqi and Dong, Ziyue and He, Jingwen and Pan, Dongwei and others},
  journal={arXiv preprint arXiv:2501.08453},
  year={2025}
}

@article{wu2024boosting,
  title={Boosting text-to-video generative model with mllms feedback},
  author={Wu, Xun and Huang, Shaohan and Wang, Guolong and Xiong, Jing and Wei, Furu},
  journal={Advances in Neural Information Processing Systems},
  volume={37},
  pages={139444--139469},
  year={2024}
}

@inproceedings{blattmann2023align,
  title={Align your latents: High-resolution video synthesis with latent diffusion models},
  author={Blattmann, Andreas and Rombach, Robin and Ling, Huan and Dockhorn, Tim and Kim, Seung Wook and Fidler, Sanja and Kreis, Karsten},
  booktitle={Proceedings of the IEEE/CVF conference on computer vision and pattern recognition},
  pages={22563--22575},
  year={2023}
}

@article{luo2023videofusion,
  title={Videofusion: Decomposed diffusion models for high-quality video generation},
  author={Luo, Zhengxiong and Chen, Dayou and Zhang, Yingya and Huang, Yan and Wang, Liang and Shen, Yujun and Zhao, Deli and Zhou, Jingren and Tan, Tieniu},
  journal={arXiv preprint arXiv:2303.08320},
  year={2023}
}

@article{sun2024sora,
  title={From sora what we can see: A survey of text-to-video generation},
  author={Sun, Rui and Zhang, Yumin and Shah, Tejal and Sun, Jiahao and Zhang, Shuoying and Li, Wenqi and Duan, Haoran and Wei, Bo and Ranjan, Rajiv},
  journal={arXiv preprint arXiv:2405.10674},
  year={2024}
}

@article{zhang2025show,
  title={Show-1: Marrying pixel and latent diffusion models for text-to-video generation},
  author={Zhang, David Junhao and Wu, Jay Zhangjie and Liu, Jia-Wei and Zhao, Rui and Ran, Lingmin and Gu, Yuchao and Gao, Difei and Shou, Mike Zheng},
  journal={International Journal of Computer Vision},
  volume={133},
  number={4},
  pages={1879--1893},
  year={2025},
  publisher={Springer}
}

@inproceedings{wu2023tune,
  title={Tune-a-video: One-shot tuning of image diffusion models for text-to-video generation},
  author={Wu, Jay Zhangjie and Ge, Yixiao and Wang, Xintao and Lei, Stan Weixian and Gu, Yuchao and Shi, Yufei and Hsu, Wynne and Shan, Ying and Qie, Xiaohu and Shou, Mike Zheng},
  booktitle={Proceedings of the IEEE/CVF international conference on computer vision},
  pages={7623--7633},
  year={2023}
}

@article{guo2023animatediff,
  title={Animatediff: Animate your personalized text-to-image diffusion models without specific tuning},
  author={Guo, Yuwei and Yang, Ceyuan and Rao, Anyi and Liang, Zhengyang and Wang, Yaohui and Qiao, Yu and Agrawala, Maneesh and Lin, Dahua and Dai, Bo},
  journal={arXiv preprint arXiv:2307.04725},
  year={2023}
}

@article{opensora2024,
  title={Open-Sora: Democratizing Efficient Video Production for All},
  author={Zheng, Zangwei and Peng, Xiangyu and Yang, Tianji and Shen, Chenhui and Li, Shenggui and Liu, Hongxin and Zhou, Yukun and Li, Tianyi and You, Yang},
   journal={URL https://github. com/hpcaitech/Open-Sora},
  year={2024}
}

@inproceedings{yuan2024instructvideo,
  title={Instructvideo: Instructing video diffusion models with human feedback},
  author={Yuan, Hangjie and Zhang, Shiwei and Wang, Xiang and Wei, Yujie and Feng, Tao and Pan, Yining and Zhang, Yingya and Liu, Ziwei and Albanie, Samuel and Ni, Dong},
  booktitle={Proceedings of the IEEE/CVF Conference on Computer Vision and Pattern Recognition},
  pages={6463--6474},
  year={2024}
}

@misc{mira2024github,
  author       = {Zhang, Zhaoyang and Yuan, Ziyang and Ju, Xuan and Gao, Yiming and Wang, Xintao and Yuan, Chun and Shan, Ying},
  title        = {Mira: A Mini-step Towards Sora-like Long Video Generation},
  year         = {2024},
  howpublished = {\url{https://github.com/mira-space/Mira}},
  note         = {Available on GitHub}
}

@inproceedings{caron2021emerging,
  title={Emerging properties in self-supervised vision transformers},
  author={Caron, Mathilde and Touvron, Hugo and Misra, Ishan and J{\'e}gou, Herv{\'e} and Mairal, Julien and Bojanowski, Piotr and Joulin, Armand},
  booktitle={Proceedings of the IEEE/CVF international conference on computer vision},
  pages={9650--9660},
  year={2021}
}

@inproceedings{radford2021learning,
  title={Learning transferable visual models from natural language supervision},
  author={Radford, Alec and Kim, Jong Wook and Hallacy, Chris and Ramesh, Aditya and Goh, Gabriel and Agarwal, Sandhini and Sastry, Girish and Askell, Amanda and Mishkin, Pamela and Clark, Jack and others},
  booktitle={International conference on machine learning},
  pages={8748--8763},
  year={2021},
  organization={PmLR}
}

@article{huang2024vbench++,
  title={Vbench++: Comprehensive and versatile benchmark suite for video generative models},
  author={Huang, Ziqi and Zhang, Fan and Xu, Xiaojie and He, Yinan and Yu, Jiashuo and Dong, Ziyue and Ma, Qianli and Chanpaisit, Nattapol and Si, Chenyang and Jiang, Yuming and others},
  journal={arXiv preprint arXiv:2411.13503},
  year={2024}
}

@inproceedings{chen2024panda,
  title={Panda-70m: Captioning 70m videos with multiple cross-modality teachers},
  author={Chen, Tsai-Shien and Siarohin, Aliaksandr and Menapace, Willi and Deyneka, Ekaterina and Chao, Hsiang-wei and Jeon, Byung Eun and Fang, Yuwei and Lee, Hsin-Ying and Ren, Jian and Yang, Ming-Hsuan and others},
  booktitle={Proceedings of the IEEE/CVF Conference on Computer Vision and Pattern Recognition},
  pages={13320--13331},
  year={2024}
}

@inproceedings{bain2021frozen,
  title={Frozen in time: A joint video and image encoder for end-to-end retrieval},
  author={Bain, Max and Nagrani, Arsha and Varol, G{\"u}l and Zisserman, Andrew},
  booktitle={Proceedings of the IEEE/CVF international conference on computer vision},
  pages={1728--1738},
  year={2021}
}

@article{wang2024vidprom,
  title={Vidprom: A million-scale real prompt-gallery dataset for text-to-video diffusion models},
  author={Wang, Wenhao and Yang, Yi},
  journal={Advances in Neural Information Processing Systems},
  volume={37},
  pages={65618--65642},
  year={2024}
}

@article{ju2024miradata,
  title={Miradata: A large-scale video dataset with long durations and structured captions},
  author={Ju, Xuan and Gao, Yiming and Zhang, Zhaoyang and Yuan, Ziyang and Wang, Xintao and Zeng, Ailing and Xiong, Yu and Xu, Qiang and Shan, Ying},
  journal={Advances in Neural Information Processing Systems},
  volume={37},
  pages={48955--48970},
  year={2024}
}

@article{miao2024t2vsafetybench,
  title={T2vsafetybench: Evaluating the safety of text-to-video generative models},
  author={Miao, Yibo and Zhu, Yifan and Yu, Lijia and Zhu, Jun and Gao, Xiao-Shan and Dong, Yinpeng},
  journal={Advances in Neural Information Processing Systems},
  volume={37},
  pages={63858--63872},
  year={2024}
}

@inproceedings{chen2020gan,
  title={Gan-leaks: A taxonomy of membership inference attacks against generative models},
  author={Chen, Dingfan and Yu, Ning and Zhang, Yang and Fritz, Mario},
  booktitle={Proceedings of the 2020 ACM SIGSAC conference on computer and communications security},
  pages={343--362},
  year={2020}
}

@article{wu2022membership,
  title={Membership inference attacks against text-to-image generation models},
  author={Wu, Yixin and Yu, Ning and Li, Zheng and Backes, Michael and Zhang, Yang},
  journal={arXiv preprint arXiv:2210.00968},
  year={2022}
}

@inproceedings{shokri2017membership,
  title={Membership inference attacks against machine learning models},
  author={Shokri, Reza and Stronati, Marco and Song, Congzheng and Shmatikov, Vitaly},
  booktitle={2017 IEEE symposium on security and privacy (SP)},
  pages={3--18},
  year={2017},
  organization={IEEE}
}

@inproceedings{li2021membership,
  title={Membership leakage in label-only exposures},
  author={Li, Zheng and Zhang, Yang},
  booktitle={Proceedings of the 2021 ACM SIGSAC Conference on Computer and Communications Security},
  pages={880--895},
  year={2021}
}

@inproceedings{lienhanced,
  title={Enhanced $\{$Label-Only$\}$ membership inference attacks with fewer queries},
  author={Li, Hao and Li, Zheng and Wu, Siyuan and Ye, Yutong and Zhang, Min and Feng, Dengguo and Zhang, Yang},
  booktitle={34th USENIX Security Symposium (USENIX Security 25)},
  pages={5465--5483},
  year={2025}
}

@inproceedings{carlini2022membership,
  title={Membership inference attacks from first principles},
  author={Carlini, Nicholas and Chien, Steve and Nasr, Milad and Song, Shuang and Terzis, Andreas and Tramer, Florian},
  booktitle={2022 IEEE symposium on security and privacy (SP)},
  pages={1897--1914},
  year={2022},
  organization={IEEE}
}

@article{truong2025attacks,
  title={Attacks and defenses for generative diffusion models: A comprehensive survey},
  author={Truong, Vu Tuan and Dang, Luan Ba and Le, Long Bao},
  journal={ACM Computing Surveys},
  volume={57},
  number={8},
  pages={1--44},
  year={2025},
  publisher={ACM New York, NY}
}

@article{maini2024llm,
  title={LLM Dataset Inference: Did you train on my dataset?},
  author={Maini, Pratyush and Jia, Hengrui and Papernot, Nicolas and Dziedzic, Adam},
  journal={Advances in Neural Information Processing Systems},
  volume={37},
  pages={124069--124092},
  year={2024}
}

@article{li2024membership,
  title={Membership inference attacks against large vision-language models},
  author={Li, Zhan and Wu, Yongtao and Chen, Yihang and Tonin, Francesco and Abad Rocamora, Elias and Cevher, Volkan},
  journal={Advances in Neural Information Processing Systems},
  volume={37},
  pages={98645--98674},
  year={2024}
}

@inproceedings{he2025towards,
  title={Towards label-only membership inference attack against pre-trained large language models},
  author={He, Yu and Li, Boheng and Liu, Liu and Ba, Zhongjie and Dong, Wei and Li, Yiming and Qin, Zhan and Ren, Kui and Chen, Chun},
  booktitle={USENIX Security},
  year={2025}
}

@article{hu2025membership,
  title={Membership inference attacks against vision-language models},
  author={Hu, Yuke and Li, Zheng and Liu, Zhihao and Zhang, Yang and Qin, Zhan and Ren, Kui and Chen, Chun},
  journal={arXiv preprint arXiv:2501.18624},
  year={2025}
}

@article{goodfellow2014generative,
  title={Generative adversarial nets},
  author={Goodfellow, Ian J and Pouget-Abadie, Jean and Mirza, Mehdi and Xu, Bing and Warde-Farley, David and Ozair, Sherjil and Courville, Aaron and Bengio, Yoshua},
  journal={Advances in neural information processing systems},
  volume={27},
  year={2014}
}

@inproceedings{arjovsky2017wasserstein,
  title={Wasserstein generative adversarial networks},
  author={Arjovsky, Martin and Chintala, Soumith and Bottou, L{\'e}on},
  booktitle={International conference on machine learning},
  pages={214--223},
  year={2017},
  organization={PMLR}
}

@article{kasneci2023chatgpt,
  title={ChatGPT for good? On opportunities and challenges of large language models for education},
  author={Kasneci, Enkelejda and Se{\ss}ler, Kathrin and K{\"u}chemann, Stefan and Bannert, Maria and Dementieva, Daryna and Fischer, Frank and Gasser, Urs and Groh, Georg and G{\"u}nnemann, Stephan and H{\"u}llermeier, Eyke and others},
  journal={Learning and individual differences},
  volume={103},
  pages={102274},
  year={2023},
  publisher={Elsevier}
}

@article{chang2024survey,
  title={A survey on evaluation of large language models},
  author={Chang, Yupeng and Wang, Xu and Wang, Jindong and Wu, Yuan and Yang, Linyi and Zhu, Kaijie and Chen, Hao and Yi, Xiaoyuan and Wang, Cunxiang and Wang, Yidong and others},
  journal={ACM transactions on intelligent systems and technology},
  volume={15},
  number={3},
  pages={1--45},
  year={2024},
  publisher={ACM New York, NY}
}

@inproceedings{momeni2023verbs,
  title={Verbs in action: Improving verb understanding in video-language models},
  author={Momeni, Liliane and Caron, Mathilde and Nagrani, Arsha and Zisserman, Andrew and Schmid, Cordelia},
  booktitle={Proceedings of the IEEE/CVF International Conference on Computer Vision},
  pages={15579--15591},
  year={2023}
}

@article{tang2025video,
  title={Video understanding with large language models: A survey},
  author={Tang, Yunlong and Bi, Jing and Xu, Siting and Song, Luchuan and Liang, Susan and Wang, Teng and Zhang, Daoan and An, Jie and Lin, Jingyang and Zhu, Rongyi and others},
  journal={IEEE Transactions on Circuits and Systems for Video Technology},
  year={2025},
  publisher={IEEE}
}

@inproceedings{menapace2024snap,
  title={Snap video: Scaled spatiotemporal transformers for text-to-video synthesis},
  author={Menapace, Willi and Siarohin, Aliaksandr and Skorokhodov, Ivan and Deyneka, Ekaterina and Chen, Tsai-Shien and Kag, Anil and Fang, Yuwei and Stoliar, Aleksei and Ricci, Elisa and Ren, Jian and others},
  booktitle={Proceedings of the IEEE/CVF Conference on Computer Vision and Pattern Recognition},
  pages={7038--7048},
  year={2024}
}

@article{liao2024evaluation,
  title={Evaluation of text-to-video generation models: A dynamics perspective},
  author={Liao, Mingxiang and Ye, Qixiang and Zuo, Wangmeng and Wan, Fang and Wang, Tianyu and Zhao, Yuzhong and Wang, Jingdong and Zhang, Xinyu and others},
  journal={Advances in Neural Information Processing Systems},
  volume={37},
  pages={109790--109816},
  year={2024}
}

@inproceedings{bar2024lumiere,
  title={Lumiere: A space-time diffusion model for video generation},
  author={Bar-Tal, Omer and Chefer, Hila and Tov, Omer and Herrmann, Charles and Paiss, Roni and Zada, Shiran and Ephrat, Ariel and Hur, Junhwa and Liu, Guanghui and Raj, Amit and others},
  booktitle={SIGGRAPH Asia 2024 Conference Papers},
  pages={1--11},
  year={2024}
}

@article{feng2024tc,
  title={Tc-bench: Benchmarking temporal compositionality in text-to-video and image-to-video generation},
  author={Feng, Weixi and Li, Jiachen and Saxon, Michael and Fu, Tsu-jui and Chen, Wenhu and Wang, William Yang},
  journal={arXiv preprint arXiv:2406.08656},
  year={2024}
}

@article{wu2023discovqa,
  title={Discovqa: Temporal distortion-content transformers for video quality assessment},
  author={Wu, Haoning and Chen, Chaofeng and Liao, Liang and Hou, Jingwen and Sun, Wenxiu and Yan, Qiong and Lin, Weisi},
  journal={IEEE Transactions on Circuits and Systems for Video Technology},
  volume={33},
  number={9},
  pages={4840--4854},
  year={2023},
  publisher={IEEE}
}

@inproceedings{teed2020raft,
  title={Raft: Recurrent all-pairs field transforms for optical flow},
  author={Teed, Zachary and Deng, Jia},
  booktitle={European conference on computer vision},
  pages={402--419},
  year={2020},
  organization={Springer}
}

@inproceedings{abadi2016deep,
  title={Deep learning with differential privacy},
  author={Abadi, Martin and Chu, Andy and Goodfellow, Ian and McMahan, H Brendan and Mironov, Ilya and Talwar, Kunal and Zhang, Li},
  booktitle={Proceedings of the 2016 ACM SIGSAC conference on computer and communications security},
  pages={308--318},
  year={2016}
}

@article{lee2021deduplicating,
  title={Deduplicating training data makes language models better},
  author={Lee, Katherine and Ippolito, Daphne and Nystrom, Andrew and Zhang, Chiyuan and Eck, Douglas and Callison-Burch, Chris and Carlini, Nicholas},
  journal={arXiv preprint arXiv:2107.06499},
  year={2021}
}

@article{wan2025wan,
title={Wan: Open and advanced large-scale video generative models},
author={Wan, Team and Wang, Ang and Ai, Baole and Wen, Bin and Mao, Chaojie and Xie, Chen-Wei and Chen, Di and Yu, Feiwu and Zhao, Haiming and Yang, Jianxiao and others},
journal={arXiv preprint arXiv:2503.20314},
year={2025}
}

@article{kong2024hunyuanvideo,
title={Hunyuanvideo: A systematic framework for large video generative models},
author={Kong, Weijie and Tian, Qi and Zhang, Zijian and Min, Rox and Dai, Zuozhuo and Zhou, Jin and Xiong, Jiangfeng and Li, Xin and Wu, Bo and Zhang, Jianwei and others},
journal={arXiv preprint arXiv:2412.03603},
year={2024}
}

@misc{GoogleAIStudio,
  author = {{Google}},
  title = {{Google AI Studio}},
  year = {2025},
  howpublished = {\url{https://aistudio.google.com/}}
}

@misc{VolcEngineDoubao,
  author = {{ByteDance}},  
  title = {{Volcano Engine: Doubao Large Model Suite}},  
  year = {2025},  
  howpublished = {\url{https://www.volcengine.com/}}, 
}

@article{zhou2022magicvideo,
  title={Magicvideo: Efficient video generation with latent diffusion models},
  author={Zhou, Daquan and Wang, Weimin and Yan, Hanshu and Lv, Weiwei and Zhu, Yizhe and Feng, Jiashi},
  journal={arXiv preprint arXiv:2211.11018},
  year={2022}
}

@misc{ffmpeg2000,
  author       = {Fabrice Bellard and FFmpeg contributors},
  title        = {FFmpeg: A Multimedia Framework},
  year         = {2000},
  howpublished = {\url{https://ffmpeg.org/}},
  note         = {Open-source software project}
}

@inproceedings{wang2024internvideo2,
  title={Internvideo2: Scaling foundation models for multimodal video understanding},
  author={Wang, Yi and Li, Kunchang and Li, Xinhao and Yu, Jiashuo and He, Yinan and Chen, Guo and Pei, Baoqi and Zheng, Rongkun and Wang, Zun and Shi, Yansong and others},
  booktitle={European Conference on Computer Vision},
  pages={396--416},
  year={2024},
  organization={Springer}
}

\cleardoublepage
\appendix
\section{Attack Algorithms}
\label{app:algorithms}
This section provides the detailed pseudo-code for the attack methodologies presented in \autoref{sec:supervised_inference}, \autoref{sec:reference_inference}, and \autoref{sec:query_inference}.

\begin{algorithm}[h!]
\caption{Supervised Inference}
\label{alg:supervised}
\begin{algorithmic}[1]
\State \textbf{Input:} Target T2V model $\mathcal{M}$, captioning model $\mathcal{C}$, labeled shadow dataset $\mathcal{D}_{\text{shadow}}$, target video $v$.
\Statex
\Procedure{Train Attack Model}{$\mathcal{D}_{\text{shadow}}, \mathcal{M}$}
    \State Initialize feature list $X \leftarrow \emptyset$, label list $Y \leftarrow \emptyset$
    \For{each $(v_i, \tilde{t}_i, y_i) \in \mathcal{D}_{\text{shadow}}$}
        \State $\mathbf{v}_{srf} \leftarrow$ \text{Compute SRF vector} $(v_i, \tilde{t}_i, \mathcal{M})$
        \State $\mathbf{s}_{instab} \leftarrow$ \text{Compute TGS vector} $(v_i, \tilde{t}_i, \mathcal{M})$
        \State $\mathbf{x}_i \leftarrow$ Concatenate ($\mathbf{v}_{srf}, \mathbf{s}_{instab}$)
        \State Append $\mathbf{x}_i$ to $X$ and $y_i$ to $Y$
    \EndFor
    \State Split $(X, Y)$ into a training set $(X_{train}, Y_{train})$ and a validation set.
    \State Train a classifier $\mathcal{A}_\theta$ on $(X_{train}, Y_{train})$.
    \State \textbf{return} Trained model $\mathcal{A}_\theta$
\EndProcedure
\Statex
\Procedure{Infer Membership}{$v, \mathcal{M}, \mathcal{A}_\theta$}
    \State $\tilde{t} \leftarrow \mathcal{C}(v)$ \text{Generate prompt}
    \State $\mathbf{v}_{srf} \leftarrow$ \text{Compute SRF vector} $(v, \tilde{t}, \mathcal{M})$
    \State $\mathbf{s}_{instab} \leftarrow$ \text{Compute TGS vector} $(v, \tilde{t}, \mathcal{M})$
    \State $\mathbf{x} \leftarrow$ Concatenate ($\mathbf{v}_{srf}, \mathbf{s}_{instab}$)
    \State $p \leftarrow \mathcal{A}_{\theta}(\mathbf{x})$
    \State \textbf{return} Membership probability $p$
\EndProcedure
\Statex
\State $\mathcal{A}_\theta \leftarrow$ Train Attack Model($\mathcal{D}_{\text{shadow}}, \mathcal{M}$)
\State $p \leftarrow$ Infer Membership($v, \mathcal{M}, \mathcal{A}_\theta$)
\State \textbf{Output:} $p$
\end{algorithmic}
\end{algorithm}

\begin{algorithm}[t!]
\caption{Reference-based Inference}
\label{alg:reference_based}
\begin{algorithmic}[1]
\State \textbf{Input:} Target T2V model $\mathcal{M}$, captioning model $\mathcal{C}$, non-member reference set $\mathcal{D}_{\text{ref}}$, target video $v$.
\Statex
\Procedure{Calibrate Statistics}{$\mathcal{D}_{\text{ref}}, \mathcal{M}$}
    \State Initialize score lists $L_{SRF} \leftarrow \emptyset$, $L_{TGS} \leftarrow \emptyset$
    \For{each $(v_i, \tilde{t}_i) \in \mathcal{D}_{ref}$}
        \State $S_{SRF, i} \leftarrow$ \text{Compute SRF Score} $(v_i, \tilde{t}_i, \mathcal{M})$ 
        \State $S_{TGS, i} \leftarrow$ \text{Compute TGS Score} $(v_i, \tilde{t}_i, \mathcal{M})$ 
        \State Append $S_{SRF, i}$ to $L_{SRF}$ and $S_{TGS, i}$ to $L_{TGS}$
    \EndFor
    \State $\mu_{srf}, \sigma_{srf} \leftarrow \text{Mean}(L_{SRF}), \text{StdDev}(L_{SRF})$
    \State $\mu_{tgs}, \sigma_{tgs} \leftarrow \text{Mean}(L_{TGS}), \text{StdDev}(L_{TGS})$
    \State \textbf{return} Statistics $(\mu_{srf}, \sigma_{srf}, \mu_{\text{tgs}}, \sigma_{\text{tgs}})$
\EndProcedure
\Statex
\Procedure{Infer Membership}{$v, \mathcal{M}, \text{stats}$}
    \State $\tilde{t} \leftarrow \mathcal{C}(v)$ \text{Generate prompt}
    \State $S_{SRF} \leftarrow$ \text{Compute SRF Score} $(v, \tilde{t}, \mathcal{M})$
    \State $S_{TGS} \leftarrow$ \text{Compute TGS Score} $(v, \tilde{t}, \mathcal{M})$
    \State $\mathcal{A}_{SRF} \leftarrow (S_{\text{SRF}} - \mu_{\text{srf}}) / \sigma_{\text{srf}}$
    \State $\mathcal{A}_{TGS} \leftarrow - (S_{\text{TGS}} - \mu_{tgs}) / \sigma_{tgs}$
    \State $\mathcal{S}_{final} \leftarrow w_{srf} \cdot \mathcal{A}_{SRF} + w_{tgs} \cdot \mathcal{A}_{TGS}$
    \State \textbf{return} Membership score $\mathcal{S}_{final}$
\EndProcedure
\Statex
\State $\text{stats} \leftarrow$ Calibrate Statistics($\mathcal{D}_{\text{ref}}, \mathcal{M}$)
\State $S \leftarrow$ Infer Membership($v, \mathcal{M}, \text{stats}$)
\State \textbf{Output:} $S$
\end{algorithmic}
\end{algorithm}

\begin{algorithm}[t!]
\caption{Query-only Inference}
\label{alg:query_only}
\begin{algorithmic}[1]
\State \textbf{Input:} Target T2V model $\mathcal{M}$, captioning model $\mathcal{C}$, target video $v$.
\Statex
\Procedure{Infer Membership}{$v, \mathcal{M}$}
    \State $\tilde{t} \leftarrow \mathcal{C}(v)$
    \State $S_{SRF} \leftarrow$ \text{Compute SRF Score} $(v, \tilde{t}, \mathcal{M})$  
    \State $S_{TGS} \leftarrow$ \text{Compute TGS Score} $(v, \tilde{t}, \mathcal{M})$  
    
    \State \text{Compute intrinsic scores: }
    \State $\mathcal{S}_{SRF} \leftarrow S_{SRF}$
    \State $\mathcal{S}_{TGS} \leftarrow 1 - S_{TGS}$
    
    \State \text{Fuse scores: }
    \State $\mathcal{S}_{final} \leftarrow w_{SRF} \cdot \mathcal{S}_{SRF} + w_{TGS} \cdot \mathcal{S}_{TGS}$
    \State \textbf{return} Membership score $\mathcal{S}_{\text{final}}$
\EndProcedure
\Statex
\State $S \leftarrow$ Infer Membership($v, \mathcal{M}$)
\State \textbf{Output:} $S$
\end{algorithmic}
\end{algorithm}

\section{Additional Robustness Analysis}
\label{app:video quality}
\begin{table}[h]
\centering
\caption{Robustness of \ourframework{} under video quality perturbations across three attack scenarios.}
\label{tab:robustness_appendix}
\resizebox{\linewidth}{!}{
\begin{tabular}{lccc}
\toprule
\textbf{Quality Setting } & \textbf{Supervised} & \textbf{Reference-based} & \textbf{Query-only} \\
\midrule
H.264 Compression & 91.70\% & 85.47\% & 81.50\% \\
Resolution Reduction & 92.52\% & 88.63\% & 82.75\% \\
\rowcolor{gray!10}
\textbf{\ourframework{}} & \textbf{93.46\%} & \textbf{88.68\%} & \textbf{82.92\%} \\
\bottomrule
\end{tabular}}
\end{table}

To further assess the robustness of \ourframework{}, we evaluate its performance under common video quality perturbations, including (1) H.264 video compression and (2) spatial resolution reduction (e.g., from $1280{\times}720$ to $480{\times}270$). 

As shown in~\autoref{tab:robustness_appendix}, \ourframework{} maintains strong AUC performance under compression and resolution reduction across all threat scenarios, indicating that \ourframework{} exploits intrinsic sparse-temporal memorization rather than low-level visual artifacts.


\end{document}